\documentclass[11pt,a4paper]{article}
\usepackage{amsmath}
\usepackage{graphicx}
\usepackage{xcolor}
\usepackage[caption=false]{subfig}
\usepackage{mathrsfs,mathtools}
\usepackage{physics,amssymb}
\usepackage{siunitx}
\usepackage{bm}
\usepackage{booktabs}
\usepackage{braket}
\usepackage{listings}
\usepackage{placeins}

\usepackage{cases}
\usepackage{comment}
\usepackage{float}
\usepackage{soul}
\usepackage{cancel}
\usepackage{cases}
\usepackage[utf8]{inputenc}
\usepackage{url}
\usepackage{float}
\usepackage{longtable}
\usepackage[normalem]{ulem}
\usepackage{xspace}
\usepackage{xcolor}
\usepackage{aas_macros}
\setstcolor{red}
\usepackage{jcappub}
\hypersetup{colorlinks=true
,urlcolor=DARKBLUE
,anchorcolor=DARKBLUE
,citecolor=DARKBLUE
,filecolor=DARKBLUE
,linkcolor=DARKBLUE
,menucolor=DARKBLUE
,linktocpage=true
,pdfproducer=medialab
}

\definecolor{MONZA}{HTML}{CF000F}
\definecolor{DARKBLUE}{HTML}{00008b}
\definecolor{DARKMAGENTA}{HTML}{8b008b}
\definecolor{DARKCYAN}{HTML}{008B8B}
\definecolor{DARKORANGE}{HTML}{FF8C00}
\definecolor{PURPLE}{HTML}{8800CC}

\begin{document}
\title{Beyond isolated curvature peaks: collective collapse and multiple Primordial Black Hole formation}
\author[a,b,c]{Albert Escriv\`a}

\affiliation[a]{Asia Pacific Center for Theoretical Physics, Pohang 37673, Republic of Korea}
\affiliation[b]{Department of Physics, Pohang University of Science and Technology, Pohang 37673, Republic of Korea}
\affiliation[c]{Department of Physics, Nagoya University, \\
Furo-cho Chikusa-ku, Nagoya 464-8602, Japan}

\emailAdd{alberto.escriva@apctp.org}
\date{\today}
\abstract{
Primordial black hole (PBH) calculations usually treat rare curvature peaks as isolated collapsing regions. Using fully nonlinear $3+1$ numerical-relativity simulations in a radiation-dominated Universe, we demonstrate that neighbouring primordial curvature perturbations need not map one-to-one onto PBHs: they may ultimately disperse, collapse collectively into a single PBH, or undergo distinct local collapses and form more than one PBH. In the family of two-component profiles studied here, the latter outcome is a pair of PBHs, identified by the coexistence of two disconnected apparent horizons on at least one time slice. We introduce a nonspherical quasi-local compaction diagnostic based on the Hawking mass and referenced to a round flat-FLRW sphere of equal area. It retains the angular structure of the curvature field and reduces to the standard Misner--Sharp compaction function in spherical symmetry. The global maximum of this diagnostic, \(\mathcal K_{\rm form}\), provides an empirical indicator of whether at least one PBH forms, with a transition near \(\mathcal K_{\rm form,c}\approx0.56\) and modest profile-dependent scatter. For forming bimodal profiles, we supplement the nonlinear compactness with a signed linear surface strength evaluated on the same probing spheres. The viability of the weaker local branch and the competition between the local and common-enclosing branches improve the empirical discrimination between single- and double-PBH outcomes. Our results therefore show that the maximum curvature amplitude alone does not determine the collapse outcome, which also depends on the spatial extent, characteristic scales, and geometry of the surrounding curvature environment.}

\maketitle
\flushbottom

\section{Introduction}

Primordial black holes (PBHs) can form from sufficiently large curvature
perturbations in the early Universe generated during inflation
\cite{Zeldovich:1967lct,10.1093/mnras/152.1.75,Carr:1974nx,Carr:1975qj}.
They are both possible dark-matter candidates and probes of primordial
fluctuations on otherwise inaccessible scales; see, e.g.,
Refs.~\cite{Sasaki:2018dmp,Carr:2020gox,Green:2020jor,Escriva:2022duf}.
Most quantitative calculations nevertheless reduce PBH formation to the
collapse of an isolated, approximately spherical high peak
\cite{1986ApJ...304...15B} for the curvature fluctuation $\zeta(\bm{x})$. However, even when $\zeta(\bm{x})$ is modeled as a Gaussian random field, its realizations contain correlated neighboring extrema and a nontrivial spatial environment \cite{1986ApJ...304...15B}. Nontrivial real-space structure has also been studied directly in
three-dimensional curvature maps generated with stochastic lattice simulations
\cite{Mizuguchi:2024kbl,Murata:2026yqb} and reconstructed nonperturbatively
from nonlinear lattice output \cite{Caravano:2024moy,Caravano:2025diq}. Complementary stochastic studies investigate spatial correlations among rare
curvature fluctuations and their associated real-space curvature and
compaction profiles \cite{Raatikainen:2023bzk,Animali:2024jiz,Raatikainen:2025gpd,Animali:2026rvi}. Full-GR simulations of post-inflationary preheating also find localized overdense regions with appreciable compactness \cite{Adshead:2023mvt}.

The abundance and spatial distribution of PBHs at formation are important
inputs for their subsequent astrophysical evolution. The initial PBH distribution and clustering have been investigated using rare-peak statistics, high-threshold correlations, excursion-set approaches, and direct realizations of correlated random fields \cite{Chisholm:2005clustering,Tada:2015biased,
AliHaimoud:2018clustering,Desjacques:2018wuu,
Suyama:2019clustering,AuclairBlachier:2024,Choi:2025eqn}.
These studies show that the predicted spatial distribution can exhibit
clustering, approximately Poisson behaviour, or short-range exclusion,
depending on the statistics and scale dependence of the primordial
fluctuations and on the assumed collapse prescription. Short-range
clustering or exclusion changes nearest-neighbour statistics and the local
tidal environment
\cite{AliHaimoud:2018clustering,Desjacques:2018wuu,
AuclairBlachier:2024}, and can thereby modify PBH binary formation,
survival, and merger rates
\cite{Clesse:2016vqa,Sasaki:2016jop,Ali-Haimoud:2017rtz,
Ballesteros:2018swv,Raidal:2018bbj,Vaskonen:2019jpv,
YoungByrnes:2019,DeLuca:2020clustering}.
The resulting merger history feeds directly into predictions for both
individually resolved gravitational-wave events and the unresolved
stochastic gravitational-wave background
\cite{Mandic:2016lcn,Wang:2016ana}. These statistical approaches generally associate sufficiently rare or
nominally supercritical regions with PBH formation, but do not resolve the
coupled nonlinear evolution that determines whether neighbouring perturbations
disperse or collapse locally or collectively. Consequently, an isolated-peak
formation threshold cannot generally be applied to each peak independently,
and the mapping from primordial peaks to PBHs need not be one-to-one.
Establishing the formation condition for the full spatial configuration,
together with its nonlinear mapping to PBHs, is therefore essential for
connecting primordial peak statistics to PBH abundance, spatial clustering,
and associated gravitational-wave signatures.

An environmental dependence of the PBH formation threshold was previously
demonstrated for concentric fluctuations in spherical symmetry in
Ref.~\cite{Escriva:2023qnq}. There, the nonlinear effect of their overlap
across different scales was shown to modify the compaction profile and lower
the critical peak compaction for PBH formation by several percent relative to the corresponding isolated fluctuations. When the two scales are sufficiently weakly coupled, the
isolated-peak limit is approximately recovered. This provided a controlled
nonlinear demonstration that the surrounding structure can alter the PBH
formation threshold. Spherical symmetry, however, forces all fluctuations
to share a common centre and therefore precludes genuine PBH multiplicity.
It leaves open the qualitatively different question of the nonlinear fate
of spatially distinct neighbouring peaks: can they collapse collectively,
remain dynamically independent, form multiple PBHs, or fail to collapse?

This issue is intrinsically nonlinear and numerical relativity is essential to address such questions. This also motivates the search for collapse thresholds and geometrical threshold estimators applicable beyond spherical symmetry. Spherical calculations have established accurate collapse thresholds for a broad class of profiles, see Ref.~\cite{Escriva:2021aeh} for a review.

By contrast, fully nonlinear nonspherical studies have so far explored a
much narrower range of configurations
\cite{Yoo:2020lmg,deJong:2021bbo,deJong:2023gsx,Yoo:2024lhp,
Escriva:2024lmm,Baumgarte:2026igz}. These studies have primarily
considered nonspherical deformations of a single, effectively isolated
curvature peak and its associated localized collapsing region. Although rare
high peaks of a Gaussian random field tend to be approximately spherical
\cite{1986ApJ...304...15B}, this statistical tendency does not imply that
nonspherical degrees of freedom are absent or dynamically irrelevant. For
instance, fully nonlinear simulations have shown that even small nonspherical
deformations can significantly delay or prevent collapse for isolated
configurations sufficiently close to the spherical critical threshold
\cite{Escriva:2024lmm}. More generally, finite-action fluctuations around
representative peaks contain radial and angular shape directions, including
higher multipoles beyond the local ellipsoidal sector
\cite{Escriva:2026dispersion}, which may significantly affect the collapse
threshold. Here we instead address a distinct source of nonsphericity:
several spatially separated curvature maxima evolve together, so that even
the number of PBHs formed need not coincide with the number of primordial
peaks. A two-component curvature profile is the simplest controlled setting
in which this spatial organization can be isolated from the more general
complexity of a random field.

To investigate this distinct problem, we evolve two neighbouring Gaussian curvature components in a radiation-dominated Universe using fully relativistic $3+1$ numerical simulations. Within the same family of initial conditions, we find three outcome classes according to apparent-horizon multiplicity: no PBH, one PBH, and two PBHs. The single-PBH class includes both collective collapse into a common horizon and configurations in which only one local region
collapses. A double-PBH outcome is identified when two disconnected
apparent horizons associated with the two local collapsing regions
coexist on at least one time slice. To characterize these outcomes without sphericalizing the initial curvature field, we construct an equal-area, FLRW-referenced compactness from the Hawking quasi-local energy, evaluated on background spheres whose centres are translated along the symmetry axis. The resulting centre- and scale-resolved compactness landscape shows that its global maximum organizes the formation of at least one PBH. For profiles containing two curvature maxima, we supplement this nonlinear
compactness with its signed linear component, evaluated on the same local and
common-enclosing sphere domains. The weaker local branch and the
local--common contrast then further organize whether one or two PBHs form. Throughout the paper, we use geometrized units with $G = c = 1$.

\section{Setup and threshold diagnostics}

The initial curvature perturbation is modelled as the superposition of two Gaussian components,
\begin{equation}
\zeta(\bm x)=
\mu_1 e^{-[(x-s)^2+y^2+z^2]/r_{m,1}^2}
+\mu_2 e^{-[(x+s)^2+y^2+z^2]/r_{m,2}^2},
\label{eq:prl_initial}
\end{equation}
where $\mu_i$ is the fluctuation amplitude and $r_{m,i}$ is the characteristic comoving length scale of the $i$th Gaussian component. The component centres are separated by $d_p=2s$, and their Gaussian standard deviations are $\sigma_i=r_{m,i}/\sqrt{2}$. For equal amplitudes and length scales, $r_{m,1}=r_{m,2}=r_m$, the curvature profile becomes bimodal when $s>r_m/\sqrt{2}$. All comoving lengths are expressed in units of a reference scale $L=1$. The profile is axisymmetric about the $x$ axis.

We typically use $H_{\rm ini}L=50$, where
$H_{\rm ini}\equiv H(t_{\rm ini})$. Most configurations use the equal-scale reference choice $r_{m,1}/L=r_{m,2}/L=\sqrt{6}/10\simeq0.245$.
To investigate the dependence on the component widths, we also consider selected unequal-scale configurations spanning
$0.175\lesssim r_{m,i}/L\lesssim0.490$, with $(\max(r_{m,1},r_{m,2}))/(\min(r_{m,1},r_{m,2}))\leq2$. This constitutes targeted sampling rather than a uniform scan of the
two-width parameter space. The spatial domain spans $x,y,z\in[-L_{\rm box},L_{\rm box}]$, with $L_{\rm box}=1.8L$. The superhorizon initial data are evolved using fully nonlinear $3+1$ numerical relativity in a radiation-dominated Universe
filled with a perfect fluid obeying $p=w\rho$, with $w=1/3$.
We define $t_{H,\rm iso}$ as the horizon-entry time of an isolated
reference Gaussian with comoving scale
$r_{m,\rm ref}=\sqrt{6}\,L/10$, such that
$a(t_{H,\rm iso})H(t_{H,\rm iso})r_{m,\rm ref}=1$.
During radiation domination,
$a(t)=a_{\rm ini}(t/t_{\rm ini})^{1/2}$ and $H(t)=1/(2t)$, giving
$t_{H,\rm iso}=a_{\rm ini}^2H_{\rm ini}r_{m,\rm ref}^2/2$
\cite{Escriva:2019nsa}. The corresponding background horizon mass is
$M_{H,\rm iso}\equiv(4\pi/3)\rho_bH^{-3}
=1/[2H(t_{H,\rm iso})]=t_{H,\rm iso}$.
For $a_{\rm ini}=1$ and $H_{\rm ini}L=50$, this gives
$t_{H,\rm iso}/L=M_{H,\rm iso}/L=1.5$.
Details of the initial-data construction, numerical setup, and validation
tests are provided in the Appendix.

For a closed two-surface $S$ with area $A[S]$ and area radius
$R_A=\sqrt{A[S]/(4\pi)}$, the Hawking energy, also commonly referred
to as the Hawking mass, is \cite{Hawking:1968}
\begin{equation}
 E_H[S]=\frac{R_A}{2}\left[
 1+\frac{1}{16\pi}\oint_S\theta_+\theta_-\,dA
 \right],
 \label{eq:prl_hawking_mass}
\end{equation}
where the null normals are normalized according to
$\ell\!\cdot\!k=-2$. We define the equal-area, FLRW-referenced
compactness by subtracting the Hawking energy of a round flat-FLRW
sphere with the same area radius,
\begin{equation}
 {\cal C}_H[S]\equiv
 \frac{2}{R_A}\left[
 E_H[S]-\frac{H_{\rm ini}^2R_A^3}{2}
 \right],
 \label{eq:prl_hawking_compactness}
\end{equation}
where $H_{\rm ini}$ is the background Hubble rate on the initial slice.

We evaluate this quantity on background coordinate spheres whose
centres are translated along the symmetry axis,
\begin{equation}
S(x_c,r):(x-x_c)^2+y^2+z^2=r^2.
\end{equation}
The probing spheres have comoving coordinate radius $r$. In the unperturbed
FLRW background, the corresponding areal radius is
$\bar R=a_{\rm ini}r$. In the perturbed geometry, the relevant radius entering
the Hawking construction is instead the area radius,
\[
R_A=a_{\rm ini}r
\left\langle e^{2\zeta}\right\rangle_\Omega^{1/2}
\left[1+O(\epsilon^2)\right],
\]
where
$\langle f\rangle_\Omega\equiv(4\pi)^{-1}\int f\,d\Omega$
denotes the coordinate solid-angle average over the probing sphere.

Representing the growing-mode initial data in comoving slicing and
using the long-wavelength solution for a perfect fluid satisfying
$p=w\rho$ gives
\begin{equation}
 {\cal C}_H(x_c,r)=
 \frac{3(1+w)}{5+3w}
 \left[
 1-\left\langle
 \left(1+r\,n^i\partial_i\zeta\right)^2
 \right\rangle_\Omega
 \right]
 +O(\epsilon^2),
 \label{eq:prl_CH_reduced}
\end{equation}
where
\begin{equation}
 \bm n=\frac{\bm x-\bm c}{r},
 \qquad
 \bm c=(x_c,0,0),
\end{equation}
and $\epsilon\sim (aH r_{m,\min})^{-1}\ll1$, with
$r_{m,\min}\equiv\min(r_{m,1},r_{m,2})$, is the long-wavelength expansion parameter controlling the gradients of both components.

For radiation domination, $w=1/3$, and the prefactor in
Eq.~\eqref{eq:prl_CH_reduced} reduces to $2/3$. It is useful to isolate the
signed contribution that is linear in the curvature field,
\begin{equation}
 \mathcal C_\ell(x_c,r)
 \equiv
 \frac{4}{3}
 \left[
  1-\left\langle
  1+r\,n^i\partial_i\zeta
  \right\rangle_\Omega
 \right]
 =
 -\frac{4}{3}r
 \left\langle n^i\partial_i\zeta\right\rangle_\Omega .
 \label{eq:prl_Cl}
\end{equation}
Directly expanding Eq.~\eqref{eq:prl_CH_reduced} gives
\begin{equation}
 \mathcal C_H
 =
 \mathcal C_\ell
 -\frac{3}{8}\mathcal C_\ell^2
 -\frac{2}{3}r^2
 {\rm Var}_\Omega\!\left(n^i\partial_i\zeta\right)
 +O(\epsilon^2),
 \label{eq:prl_CH_linear}
\end{equation}
where
${\rm Var}_\Omega(f)\equiv
\langle f^2\rangle_\Omega-\langle f\rangle_\Omega^2$.
The quantity $\mathcal C_\ell$ is the linear compaction of the angularly
averaged curvature profile about the chosen trial centre; it is therefore a
signed first-moment companion to $\mathcal C_H$, not a second quasi-local
mass. In spherical symmetry,
$\mathcal C_\ell=-4r\zeta'/3$ and the variance term vanishes, recovering
$\mathcal C_H=\mathcal C_\ell-3\mathcal C_\ell^2/8$ and the standard
background-subtracted Misner--Sharp compaction
\cite{Shibata:1999zs,Harada:2015yda}.
The complete derivation and gauge conventions are given in the Appendix.

\section{Nonlinear gravitational collapse}
\label{sec:non_linear_collapse}

\begin{figure*}[t]
\centering
\includegraphics[width=1.0\textwidth]{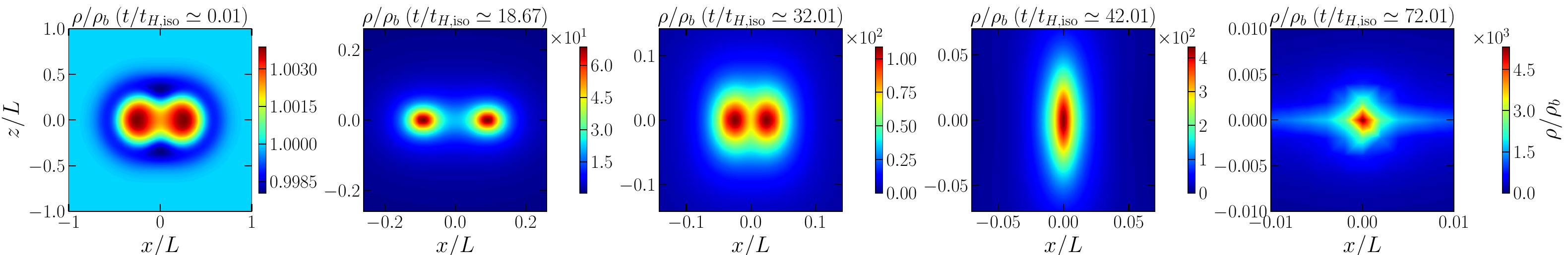}
\includegraphics[width=0.31\textwidth]{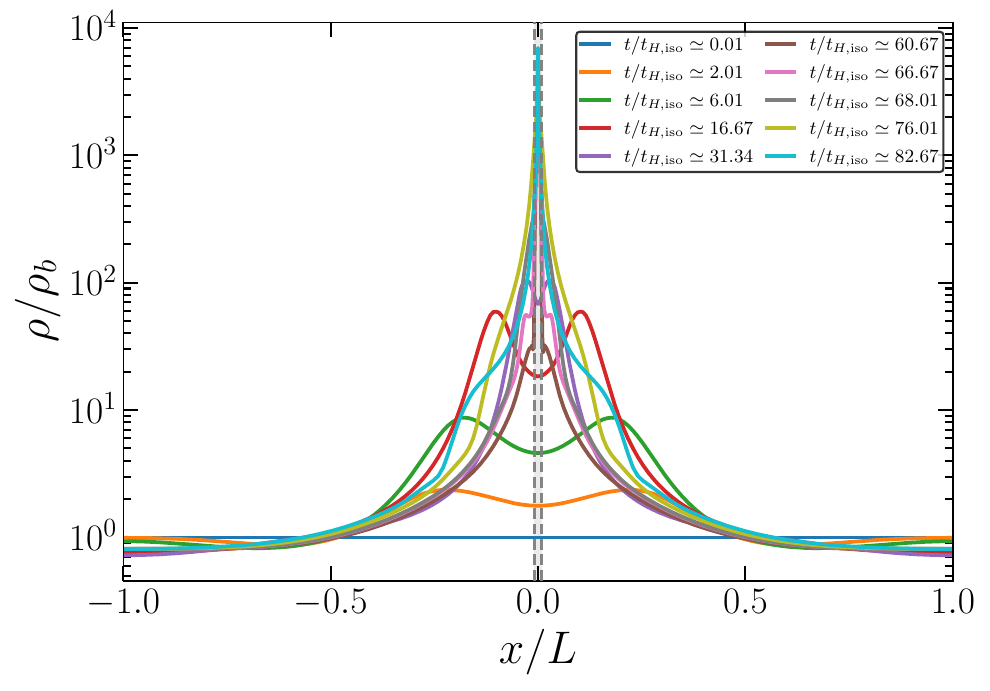}
\includegraphics[width=0.31\textwidth]{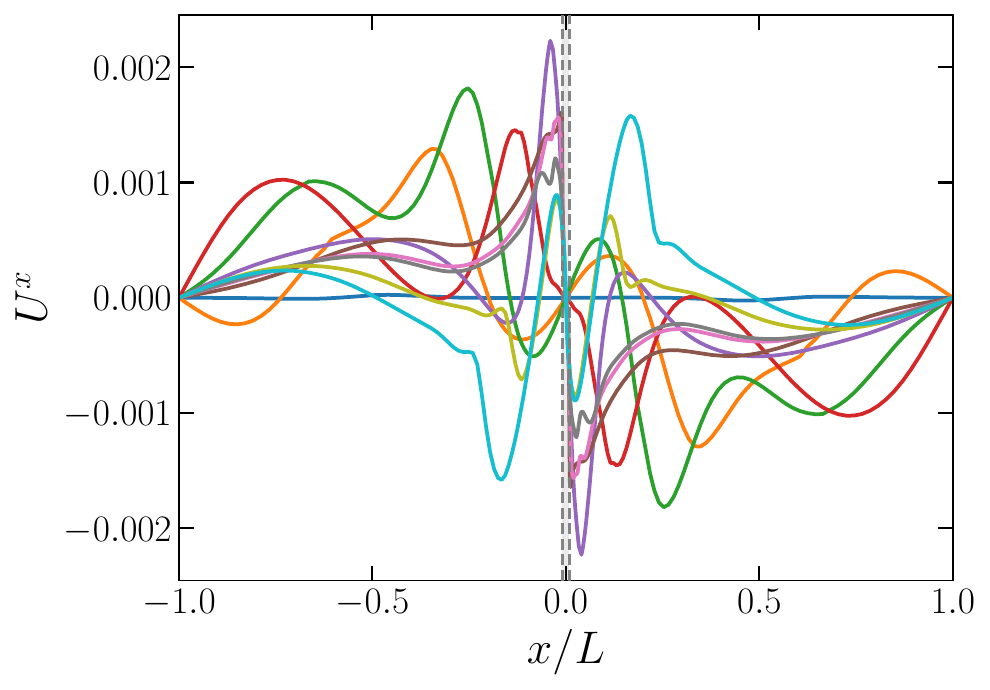}
\includegraphics[width=0.31\textwidth]{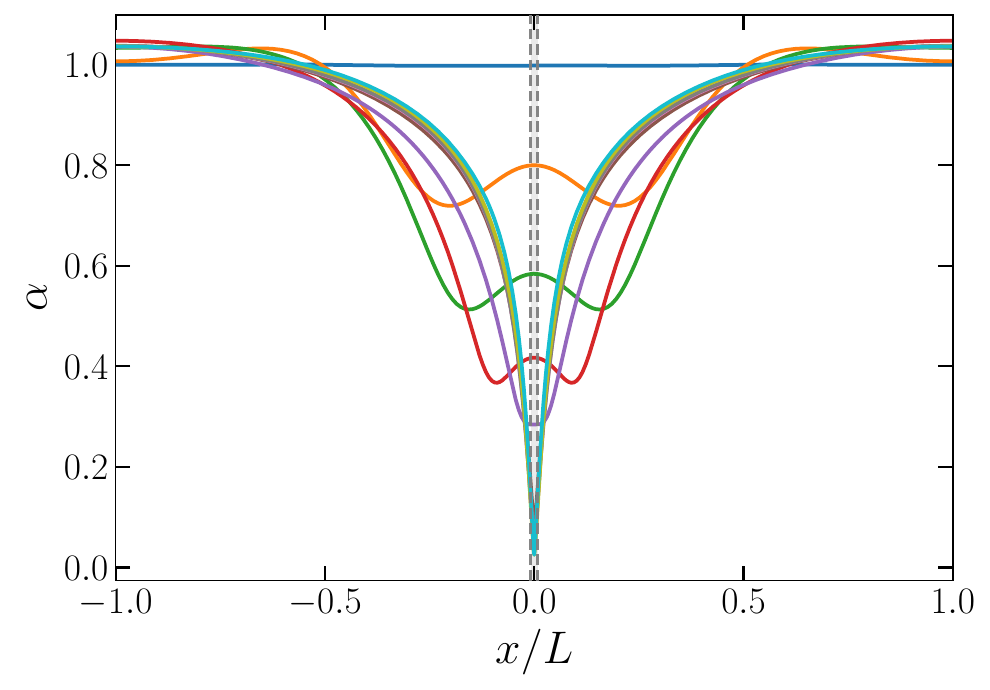}
\caption{Nonlinear evolution of the marginal collective single-PBH
configuration $(\mu_1,\mu_2,s)=(0.775,0.775,0.2)$, with
$r_{m,1}/L=r_{m,2}/L=\sqrt{6}/10\simeq0.245$ and $H_{\rm ini}L=50$.
The upper row shows $\rho/\rho_b$ in the $z=0$ plane, while the lower
panels show axial cuts of the density, fluid velocity $U^x$, and lapse
$\alpha$ at selected times. The grey shaded band bounded by dashed vertical lines in the lower panels indicates the $x$-coordinate extent of the first detected common apparent horizon, $|x_{\rm AH}|/L\simeq0.00876$, at $t/t_{H,\rm iso}\simeq72.28$.}
\label{fig:prl_single_dynamics1}
\end{figure*}

Across our numerical sample, we find non-forming, single-PBH, and double-PBH outcomes. Here we highlight three representative evolutions, with further diagnostics presented in the Appendix. Sufficiently close Gaussian components can collapse collectively even when
neither would be supercritical in isolation. For example, the configuration
$(\mu_1,\mu_2,s)=(0.775,0.775,0.2)$ has two maxima in the total initial
curvature profile but develops a common apparent horizon enclosing both
regions. It is therefore classified as a single-PBH outcome.
Figure~\ref{fig:prl_single_dynamics1} shows its nonlinear evolution.

At the initial time, the density field displays two distinct maxima associated with the two curvature components. As the perturbation enters the nonlinear regime, both overdense regions contract and a convergent fluid flow develops between them. This convergence is indicated by the opposite signs of the Eulerian velocity $U^x$ (defined following Ref.~\cite{Escriva:2024lmm}) on the two sides of the origin. The two density concentrations
progressively connect and evolve into a single centrally concentrated
overdensity with sharp gradients. At the same time, the density contrast grows by several orders of magnitude and the lapse develops an increasingly deep minimum
around the common centre, consistently with the onset of strong-field
collapse in the adopted slicing. The evolution therefore does not proceed as two independent collapses. Instead, the initially two-centred configuration dynamically assembles a common collapsing region that ultimately forms a single PBH. Here $\mu_1=\mu_2=0.775$ lies below the isolated-Gaussian formation threshold
$\mu_c^{\rm iso}\simeq0.796$.

\begin{figure}[t]
\centering
\includegraphics[width=0.5\textwidth]{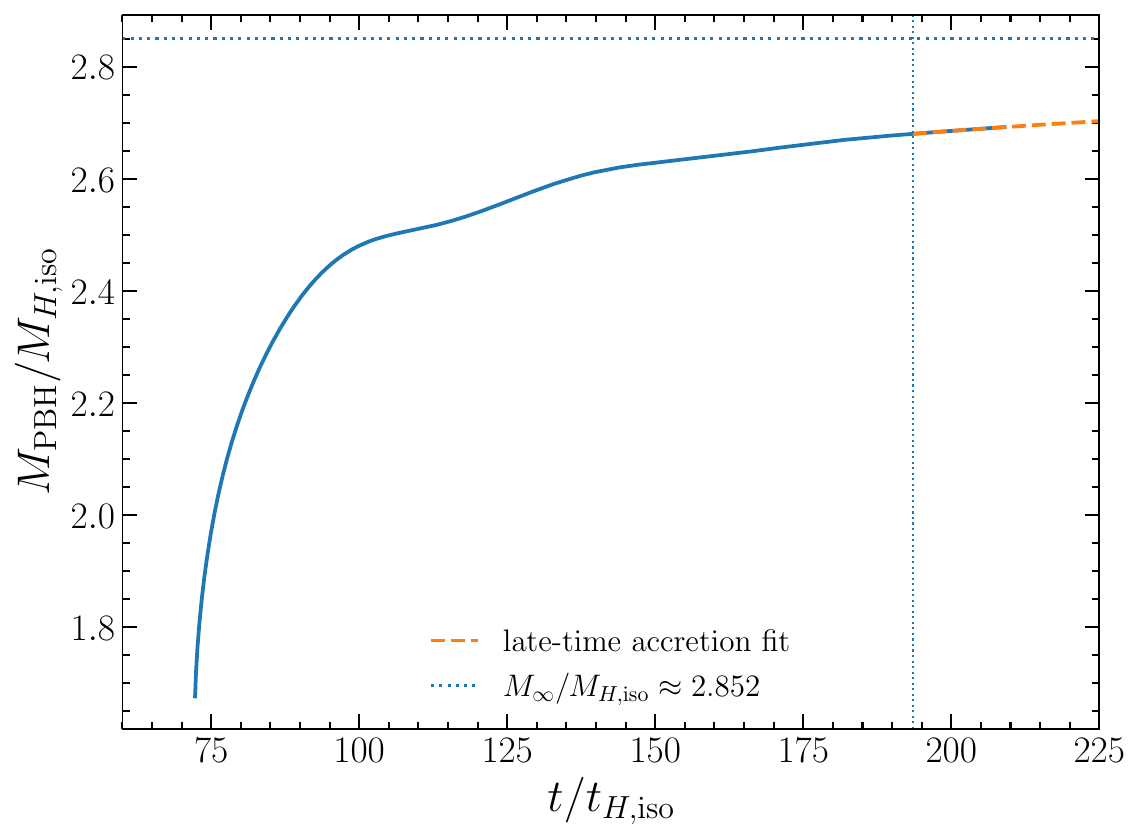}
\caption{Evolution of the apparent-horizon mass for the configuration
shown in Fig.~\ref{fig:prl_single_dynamics1}, normalized by the horizon
mass $M_{H,\rm iso}$ associated with the characteristic scale of a single
Gaussian component. The dashed curve is a late-time fit to the
Zel'dovich--Novikov accretion model. Its extrapolated asymptotic
value is $M_\infty/M_{H,\rm iso}\simeq2.85$.}
\label{fig:pbh_mass_evolution}
\end{figure}

To quantify how collective collapse modifies the mass of the resulting PBH relative to the isolated-component expectation, we follow the apparent-horizon mass after formation. Its evolution is shown in Fig.~\ref{fig:pbh_mass_evolution}, displaying the typical behaviour of an initial increase due to accretion from the surrounding cosmological fluid, followed by an approach to an asymptotic value. Here $M_{H,\rm iso}$ denotes the horizon mass associated with the characteristic scale of a single isolated Gaussian component. Because the common collapsing region encompasses both components, it is associated with an effective mass larger than that of either component separately. A fit to the late-time mass evolution using the Zel'dovich--Novikov accretion model \cite{Zeldovich:1967lct,Guedens:2002sd,Nayak:2009wk},
following the approach of Refs.~\cite{Escriva:2019nsa,Escriva:2025eqc},
extrapolates to $M_\infty/M_{H,\rm iso}\simeq 2.85$.

\begin{figure*}[t]
\centering
\includegraphics[width=1.0\textwidth]{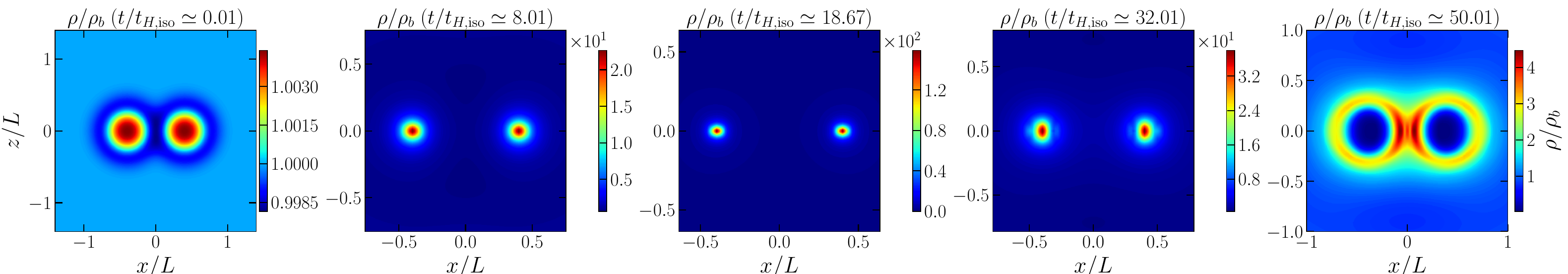}
\includegraphics[width=0.31\textwidth]{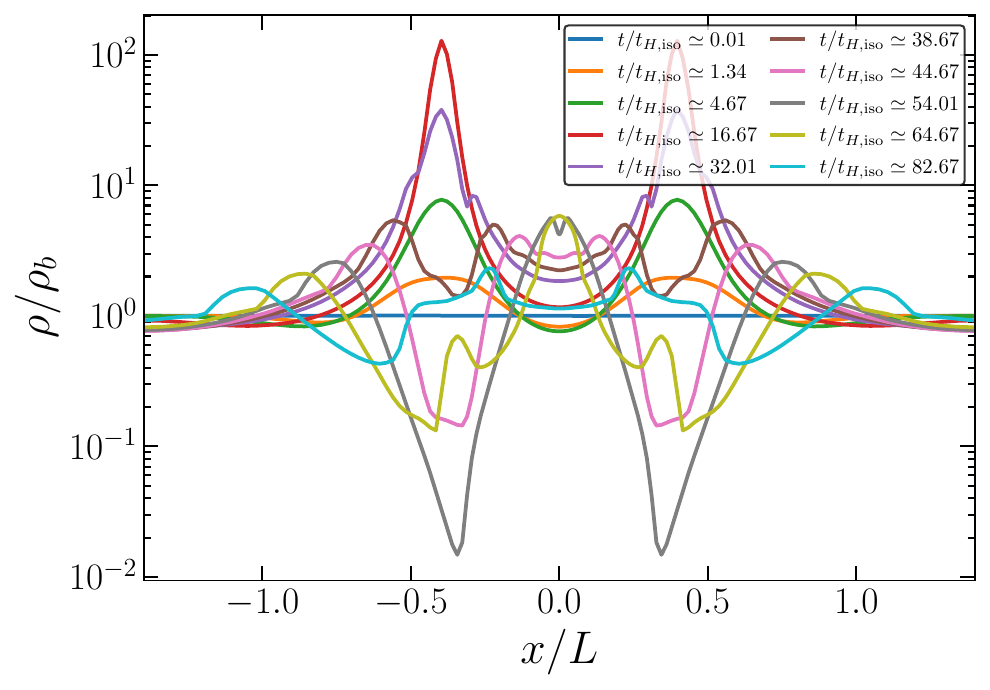}
\includegraphics[width=0.31\textwidth]{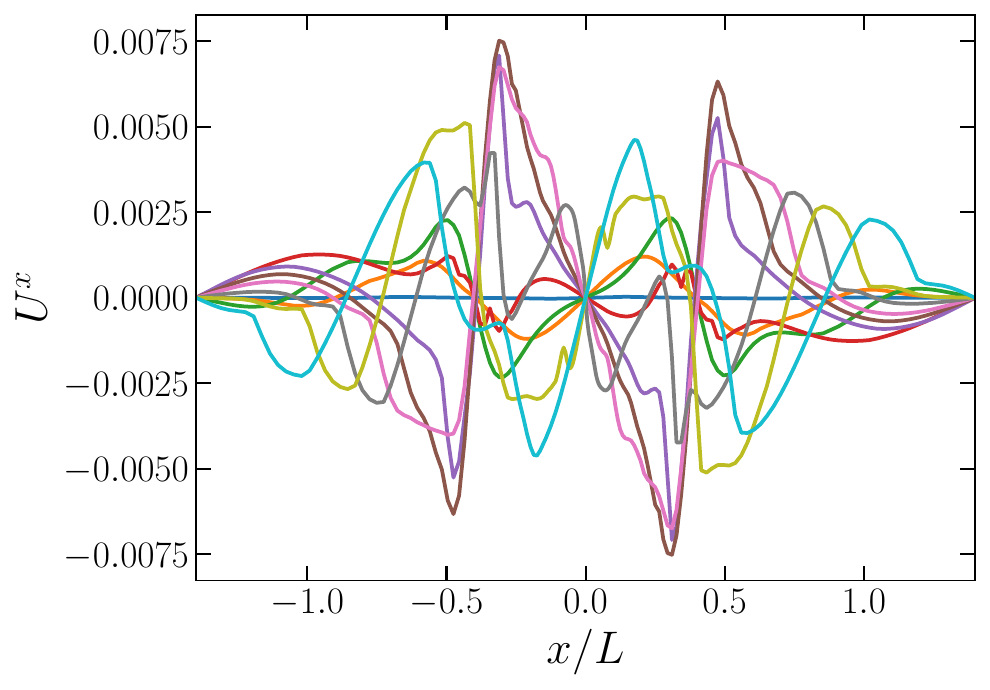}
\includegraphics[width=0.31\textwidth]{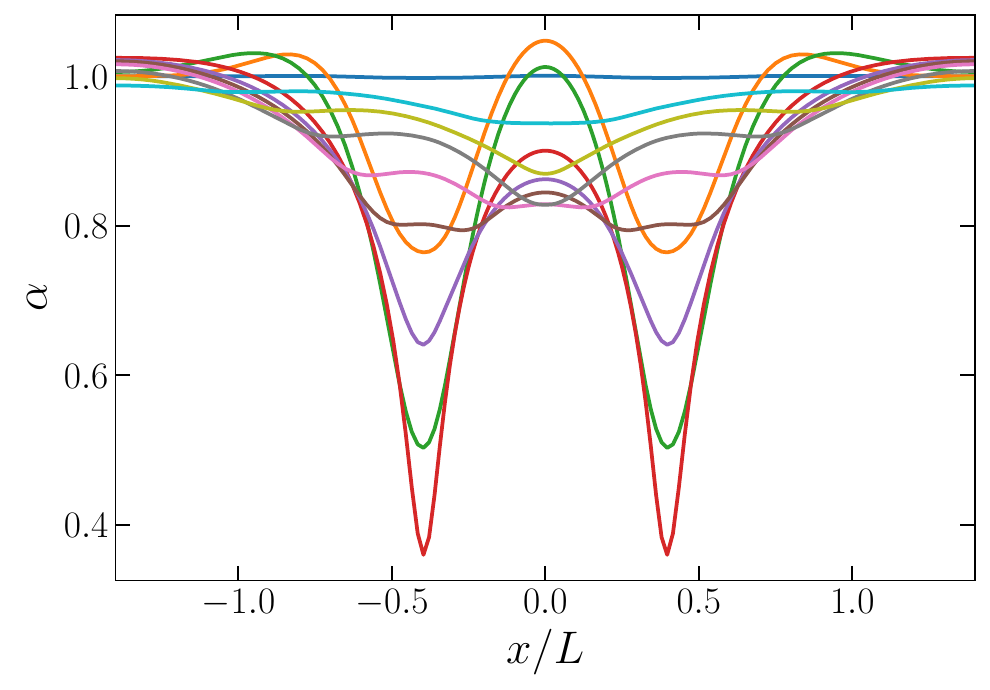}
\caption{Same dynamical diagnostics as in
Fig.~\ref{fig:prl_single_dynamics1}, for the separated non-forming
configuration $(\mu_1,\mu_2,s)=(0.775,0.775,0.4)$.}
\label{fig:prl_no_dynamics}
\end{figure*}

Increasing the separation can prevent collective collapse. For
$(\mu_1,\mu_2,s)=(0.775,0.775,0.4)$, the two overdensities remain
spatially distinct and no apparent horizon is detected during the
evolution, as shown in Fig.~\ref{fig:prl_no_dynamics}. Both regions
undergo strong nonlinear compression and develop distinct lapse minima,
but no persistent convergent flow towards a common centre is established.
At later times, the velocity reverses, the density peaks disperse into
outward-propagating shells, and the lapse tends to recover its background value. At this larger
separation, pressure gradients are therefore sufficient to halt the
collapse and drive re-expansion before a common trapped region can form.

\begin{figure*}[t]
\centering
\includegraphics[width=1.0\textwidth]{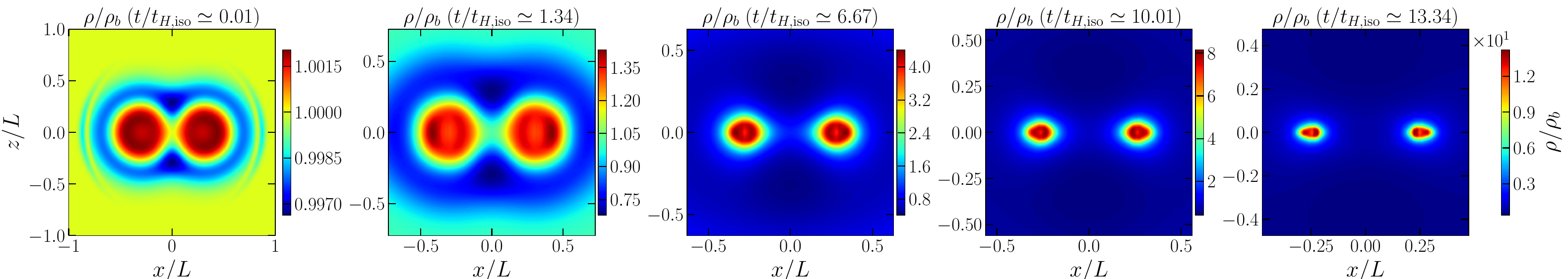}
\includegraphics[width=0.31\textwidth]{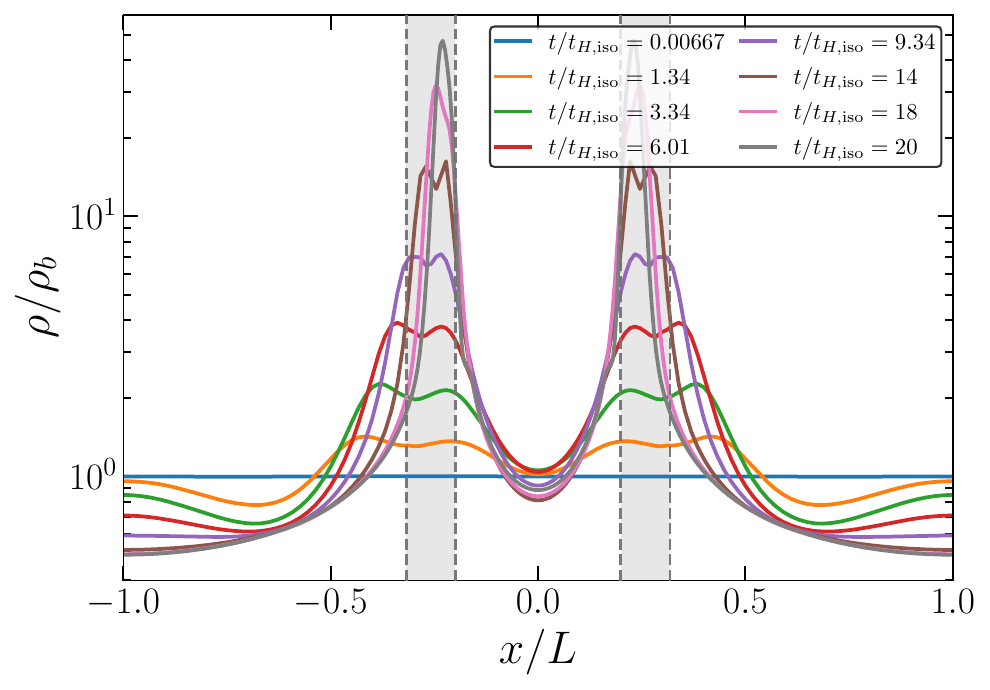}
\includegraphics[width=0.31\textwidth]{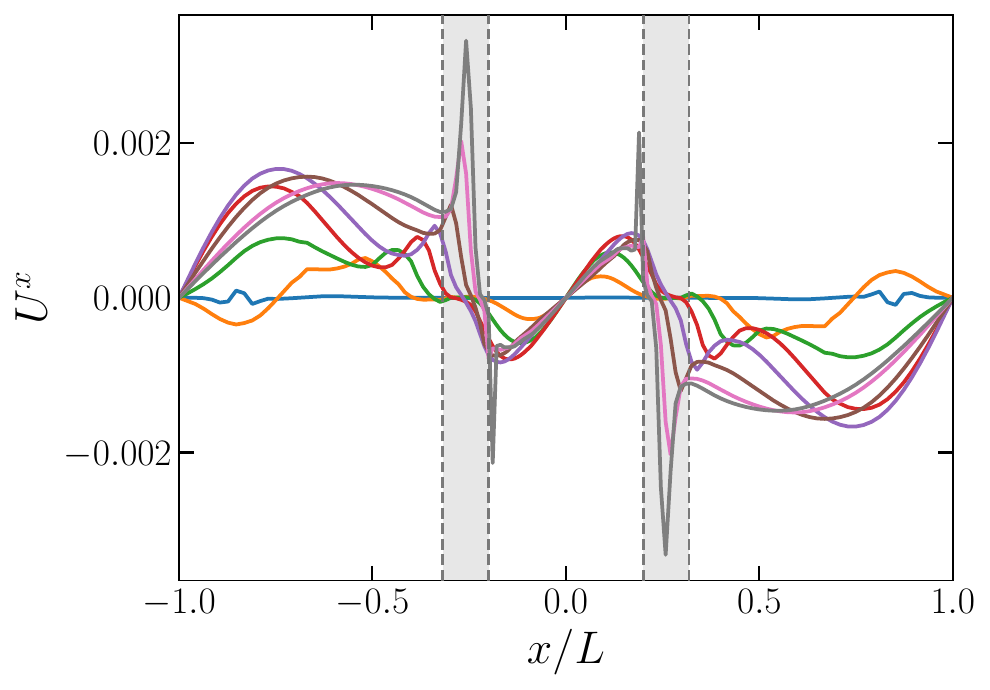}
\includegraphics[width=0.31\textwidth]{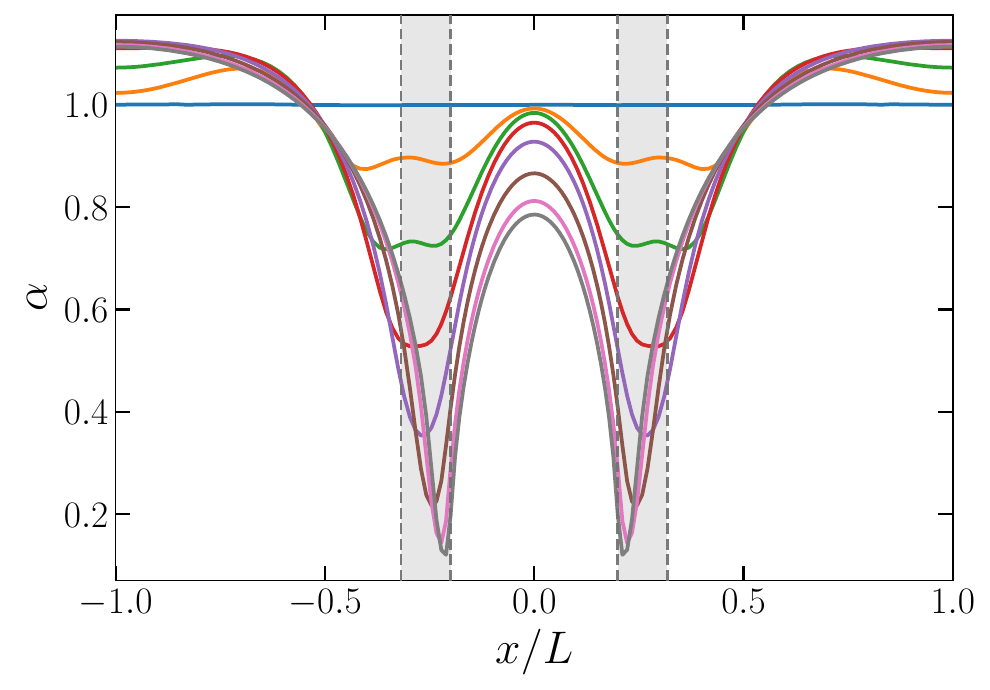}
\caption{Same dynamical diagnostics as in
Fig.~\ref{fig:prl_single_dynamics1}, for a representative strong
double-PBH configuration with
$(\mu_1,\mu_2,s)=(1.6,1.6,0.3)$. The grey shaded bands mark the first detected pair of local apparent
horizons, with coordinate extents $-0.318\lesssim x/L\lesssim-0.200$ and
$0.200\lesssim x/L\lesssim0.318$, at
$t/t_{H,\mathrm{iso}}\simeq15.22$.}
\label{fig:prl_double_dynamics}
\end{figure*}

The third outcome is genuine multiplicity. Figure~\ref{fig:prl_double_dynamics} shows the
nonlinear evolution of a representative double-PBH configuration with
$(\mu_1,\mu_2,s)=(1.6,1.6,0.3)$. The two overdense regions remain
spatially distinct as each contracts around its own centre. Their density
maxima grow rapidly and become increasingly narrow without developing a
comparably dense bridge. Correspondingly, the velocity field shows
convergent flows towards the individual overdensities rather than towards
the midpoint. The lapse develops two deep, well-separated minima,
indicating two distinct strong-field regions in the adopted slicing. The collapsing regions therefore remain dynamically distinct up to
black-hole formation. The apparent-horizon finder detects two disconnected local apparent
horizons, signalling the formation of two PBHs. Within the symmetry setting of our simulations, these double-PBH configurations may provide a primordial channel for head-on PBH mergers.
If the pair remains bound, a common apparent horizon may later form around
the two local horizons, providing a quasi-local signature of merger; the associated collision would emit gravitational waves
\cite{Anninos:1993zj,Sperhake:2005uf}. Such primordial formation-era mergers could
constitute an additional classical source of gravitational waves in the
early Universe. For $\mu_1=\mu_2=1.6$, changing only the half-separation from $s=0.2$ to
$s=0.3$ changes the outcome from collective single-PBH collapse to the
formation of two PBHs; the corresponding initial compactness landscapes
are compared in Fig.~\ref{fig:figS_hawking_basin_landscapes.pdf}. These examples show that PBH multiplicity depends on the spatial organization of neighbouring perturbations and motivate diagnostics that distinguish local from common collapse.

\section{Quasi-local organization of collapse}
\label{sec:hawking_organization}

\begin{figure*}[!t]
\centering
\includegraphics[width=0.96\textwidth]
{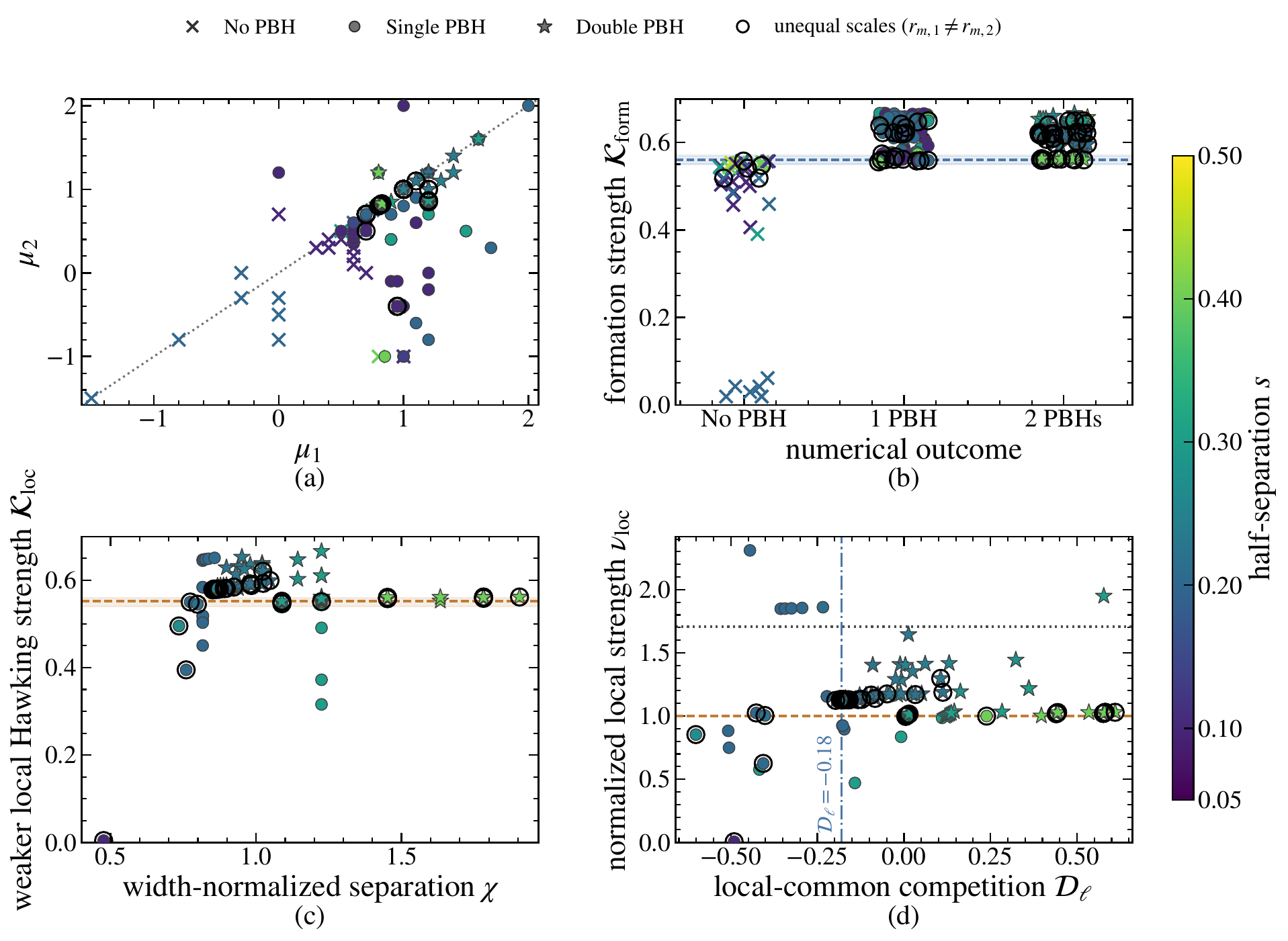}
\caption{Collapse organization for two-component curvature profiles across the simulated cases. Marker shape indicates the numerical outcome, colour encodes the
half-separation $s$, and black rings mark unequal component scales.
(a) Numerical outcomes in the $(\mu_1,\mu_2)$ plane.
(b) Global Hawking formation strength $\mathcal K_{\rm form}$ for the
complete numerical sample. The shaded $0.55$--$0.56$ band indicates the
empirical formation-transition region, and the dashed line marks
$\mathcal K_{\rm form}=0.56$.
(c) Nonlinear local--separation organization for PBH-forming profiles
containing two curvature maxima. The vertical coordinate $\mathcal K_{\mathrm{loc}}$
is the weaker local Hawking strength, while $\chi$ measures the
width-normalized component separation. The shaded $0.54$--$0.56$ band
is a conservative local-strength reference, and the dashed line marks
the isolated-Gaussian value $\mathcal K_{\rm iso}\simeq0.552$. The
compression of $\mathcal K_{\mathrm{loc}}$ near its nonlinear turnover produces
appreciable overlap between the single- and double-PBH outcomes.
(d) Signed linear-branch organization for the same profiles.
The normalized weaker-local strength $\nu_{\rm loc}$ measures local
viability, while $\mathcal D_\ell$ measures local--common competition.
The horizontal dashed line marks the isolated-Gaussian reference
$\nu_{\rm loc}=1$, the horizontal dotted line marks the spherical
Type-I/Type-II reference $\nu_{\rm II}\simeq1.71$, and the vertical
dash-dotted line at $\mathcal D_\ell=-0.18$ is an empirical
multiplicity guide. Most double-PBH outcomes lie above $\nu_{\rm loc}=1$ and to the right
of $\mathcal D_\ell=-0.18$, but the remaining overlap shows that these
guides do not define a universal separator.}
\label{fig:prl_classification}
\end{figure*}

We now examine which initial-data quantities organize the nonlinear outcomes. Panel~(a) of Fig.~\ref{fig:prl_classification} shows that the component
amplitudes $\mu_1$ and $\mu_2$ alone are insufficient: collective collapse
can occur when both lie below $\mu_c^{\rm iso}$, whereas some configurations
remain non-forming even when both are close to or slightly above this
isolated-Gaussian threshold.

The global maximum of the total curvature field,
$\zeta_{\rm pk}\equiv\max_{\bm x}\zeta(\bm x)$, does not remove this
ambiguity. Forming configurations extend down to
$\zeta_{\rm pk}=0.768$, while non-forming configurations extend up to
$\zeta_{\rm pk}=0.820$. In particular, the configurations
$(\mu_1,\mu_2,s;r_{m,1}/L,r_{m,2}/L)
=(0.7,0.7,0.2;0.245,0.280)$ and
$(0.82,0.82,0.4;0.245,0.175)$
have nearly identical values,
$\zeta_{\rm pk}=0.8196$ and $0.8200$, yet produce no PBH and two PBHs,
respectively. Among forming configurations, $\zeta_{\rm pk}$ also does not distinguish single- from double-PBH formation. Because $\zeta_{\rm pk}$ is pointwise, it does not capture the spatial and angular structure that controls the quasi-local mass excess and pressure support relevant for collapse. In spherical symmetry, formation thresholds are instead commonly characterized using the compaction function (see Ref.~\cite{Escriva:2021aeh} for a review), which measures the mass excess over a range of scales. This motivates a centre- and scale-resolved nonspherical extension based on the FLRW-referenced Hawking compactness. Scanning the Hawking compactness over radius and over centres $\bm c=(x_c,0,0)$ translated along the peak axis, we define the global formation strength

\begin{equation}
 \mathcal K_{\rm form}
 =
 \max_{x_c,r}\mathcal C_H(x_c,r).
 \label{eq:prl_Kform}
\end{equation}
The maximization is performed over the whole domain. Since $\mathcal C_H$ measures a dimensionless background-subtracted
quasi-local energy, $\mathcal K_{\rm form}$ selects the strongest compact
region over all probed centres and scales. It is defined for both unimodal
and bimodal profiles and provides the cleanest one-number organization of
whether at least one PBH forms. Across the numerical sample, the transition occurs near the empirical
reference $\mathcal K_{\rm form,c}\simeq0.56$. The residual spread,
$0.55\lesssim\mathcal K_{\rm form}\lesssim0.56$, is interpreted as a
profile-dependent transition region rather than a sharp universal threshold,
consistent with the profile dependence found in spherical-collapse simulations
\cite{Escriva:2019phb}.

Multiplicity requires spatial information beyond the global maximum.
For profiles with two curvature maxima at
$x_{\zeta,L}<x_{\zeta,R}$, let $x_s$ denote the intervening minimum of
$\zeta(x,0,0)$. We restrict local probing spheres to remain on one side
of the corresponding saddle plane:
\begin{align}
 \mathcal B_L
 &=
 \{(x_c,r):x_c<x_s,\ r<x_s-x_c\},
 \\
 \mathcal B_R
 &=
 \{(x_c,r):x_c>x_s,\ r<x_c-x_s\}.
 \label{eq:prl_local_basins}
\end{align}
Each basin defines a local-branch strength,
\begin{equation}
\mathcal K_{\mathrm{loc}}^{(i)}
\equiv
\max_{(x_c,r)\in\mathcal B_i}
\mathcal C_H(x_c,r),
\qquad
\mathcal K_{\mathrm{loc}}
\equiv
\min_{i\in\{L,R\}}
\mathcal K_{\mathrm{loc}}^{(i)}.
 \label{eq:prl_KL}
\end{equation}
Thus $\mathcal K_{\mathrm{loc}}$ is the weaker of the two basin-restricted local
maxima. The minimum is taken because the formation of two PBHs requires
both local regions to be sufficiently strong: the weaker branch therefore
acts as the bottleneck for multiplicity. The quantity $\mathcal K_{\mathrm{loc}}$ is
defined only for bimodal curvature profiles, since its construction requires
two distinct local branches.

For comparison, an isolated spherical Gaussian reaches the formation
threshold at $\mu_c^{\rm iso}\simeq0.796$, corresponding to $\mathcal K_{\rm iso}
\equiv
\max_r\mathcal C_H(0,r)
\simeq0.552$. Since $\mathcal C_H$ coincides with the standard
background-subtracted Misner--Sharp compaction in spherical symmetry,
$\mathcal K_{\rm iso}$ is the corresponding Gaussian compaction
threshold. For bimodal profiles, it provides a reference for the
strength of each local branch. We show
$0.54\lesssim\mathcal K_{\mathrm{loc}}\lesssim0.56$ as a conservative reference
interval, not as an exact or necessary condition for local collapse,
since the threshold remains profile dependent.

As an input measure of geometrical overlap, we define the
width-normalized separation
\begin{equation}
\chi\equiv\frac{2s}{r_{m,1}+r_{m,2}}.
\label{eq:prl_chik}
\end{equation}
Small $\chi$ corresponds to strongly overlapping components, whereas
larger $\chi$ describes components that are more widely separated
relative to their characteristic widths.

For PBH-forming bimodal profiles, the pair
$(\chi,\mathcal K_{\mathrm{loc}})$ provides a physically intuitive first
organization of multiplicity. The quantity $\mathcal K_{\mathrm{loc}}$ tests the
viability of the weaker local region, while $\chi$ measures whether the
two components are sufficiently distinct relative to their widths.
Two-PBH formation is favoured when both local regions are viable and
geometrically distinct, whereas collective single-PBH collapse is
favoured when a common-enclosing mode dominates.

However, Eq.~\eqref{eq:prl_CH_linear} shows that the nonlinear Hawking compactness is quadratic in the signed linear strength and also contains a negative angular-variance contribution. In spherical symmetry, the quadratic relation turns over at $\mathcal C_\ell=4/3$, where $\mathcal C_H=2/3$. This value separates Type-I profiles, for which the areal radius is monotonic, from Type-II profiles, for which it is nonmonotonic \cite{Kopp:2010sh}. We use this value only as a spherical reference for the nonspherical local branches considered here. As their linear strength approaches or exceeds this value, $\mathcal K_{\mathrm{loc}}$ becomes compressed and no longer orders strong local branches monotonically \cite{Escriva:2025rja}. The angular-variance term introduces an additional source of degeneracy. This behaviour is visible in Fig.~\ref{fig:prl_classification}(c) and motivates evaluating $\mathcal C_\ell$ on the same local and common-enclosing domains. For the signed linear diagnostic, let $\mathcal B_C$ denote the
common-enclosing domain of spheres containing both curvature maxima,
whose explicit construction is given in the Appendix. For $i=L,R$, we define
\begin{align}
\mathcal K_{\ell,\mathrm{loc}}^{(i)}
&\equiv
\max_{(x_c,r)\in\mathcal B_i}
\mathcal C_\ell(x_c,r),
\\
\mathcal K_{\ell,\mathrm{loc}}
&\equiv
\min_{i\in\{L,R\}}
\mathcal K_{\ell,\mathrm{loc}}^{(i)},
\\
\mathcal K_{\ell,C}
&\equiv
\max_{(x_c,r)\in\mathcal B_C}
\mathcal C_\ell(x_c,r).
 \label{eq:prl_Kell_branches}
\end{align}
For an isolated Gaussian at its formation threshold,
$\mathcal K_{\ell,\rm iso}\equiv
\max_r\mathcal C_\ell(0,r)\simeq0.781$. We then introduce
\begin{equation}
\nu_{\mathrm{loc}}
\equiv
\frac{\mathcal K_{\ell,\mathrm{loc}}}
     {\mathcal K_{\ell,\mathrm{iso}}},
\qquad
\mathcal D_\ell
\equiv
\frac{\mathcal K_{\ell,\mathrm{loc}}-\mathcal K_{\ell,C}}
     {\mathcal K_{\ell,\mathrm{iso}}}.
     \label{eq:prl_nu_Dell}
\end{equation}
Here $\nu_{\rm loc}$ measures the viability of the weaker local branch,
whereas $\mathcal D_\ell$ measures local--common competition: positive
values indicate local dominance and negative values common-enclosing
dominance. The spherical Type-I/Type-II reference
$\mathcal C_\ell=4/3$ corresponds to
$\nu_{\rm II}\equiv
(4/3)/\mathcal K_{\ell,\rm iso}\simeq1.71$.

The PBH multiplicity is determined directly from the numerical
apparent-horizon structure. A configuration is classified as a
double-PBH outcome when two spatially disconnected apparent horizons
associated with the two local collapsing regions coexist on at least
one time slice, irrespective of whether they form simultaneously or
sequentially. The single-PBH category is likewise defined by apparent-horizon multiplicity rather than by the collapse mechanism. It includes both
collective collapse into a common horizon and configurations in which only one local region forms an apparent horizon while the other disperses. These two channels are illustrated, respectively, in
Figs.~\ref{fig:prl_single_dynamics1} and
\ref{fig:prl_single_dynamics55}.

The results suggest a simple physical interpretation. The global strength $\mathcal K_{\rm form}$ tests whether the initial data contain any sufficiently compact region to form a PBH. Once a forming bimodal configuration is selected, the multiplicity reflects a competition between the separate collapse of the two local regions and the collapse of a larger region enclosing both of them. The quantity $\nu_{\rm loc}$ tests the viability of the weaker local region, which acts as the bottleneck for forming two PBHs, while $\mathcal D_\ell$ compares this weaker local branch with the common-enclosing branch. A double-PBH outcome therefore requires both local regions to be viable and the common-enclosing branch not to dominate too strongly. By contrast, a single PBH can result either because one local region is not sufficiently strong or because the common region collapses collectively.

Across the present sample, most double-PBH outcomes satisfy $\nu_{\rm loc}\gtrsim1$ and $\mathcal D_\ell\gtrsim-0.18$, whereas most single-PBH outcomes fail at least one of these conditions. The input geometry influences this competition. The separation $\chi$ measures the overlap of the two components, while the component-scale ratio controls their relative horizon-entry times. In radiation domination, $t_{H,i}\propto r_{m,i}^{2}$, so unequal scales can cause the two regions to enter the horizon and collapse at different times. Strong overlap tends to favour a common collapsing region, whereas greater separation allows the local regions to evolve more independently.

These trends provide an empirical organization of the numerical outcomes rather than universal thresholds. The Hawking compactness remains the quasi-local formation diagnostic, including the negative angular-variance contribution in Eq.~\eqref{eq:prl_CH_linear}. The signed quantity $\mathcal C_\ell$ is used only to compare the local and common-enclosing branches and should not be interpreted as a second quasi-local mass. 


\clearpage
\section{Conclusions}

Our simulations demonstrate that the isolated-peak approximation can
fail even for a minimal configuration consisting of two neighboring
Gaussian curvature components. Within the parameter range studied,
the perturbations may disperse, form a single PBH, or form two PBHs
with separate horizons. The single-PBH outcomes include both collective
collapse into a common horizon and configurations in which only one
local region collapses. These outcomes cannot be determined from the
component amplitudes or their separation alone. Instead, PBH formation
and multiplicity depend on the overall strength, component widths,
separation, and geometry of the initial perturbation, which determine
whether collapse proceeds locally or collectively.

The equal-area, FLRW-referenced Hawking compactness provides a quasi-local extension of the standard PBH compaction function to nonspherical settings without sphericalizing the initial curvature field. It retains the angular dependence and reduces to the Misner--Sharp
compaction in spherical symmetry. This construction provides the first estimate of the PBH formation threshold based on a genuinely nonspherical compaction function. For the two-component family studied here, the global strength $\mathcal K_{\rm form}$ organizes the formation of at least one PBH around an empirical reference value of $0.56$, with the interval $0.55\lesssim\mathcal K_{\rm form}\lesssim0.56$ representing the observed profile-dependent transition region in the present numerical sample.

For profiles containing two curvature maxima, the multiplicity reflects a competition between the viability of both local collapsing regions and the tendency of the surrounding configuration to collapse collectively. The signed linear branch quantities $\nu_{\rm loc}$ and $\mathcal D_\ell$ provide initial-data proxies for these two effects and improve the empirical organization of the single- and double-PBH outcomes. The remaining overlap, however, shows that they do not define an exact multiplicity criterion. The separation and component-scale ratio remain relevant secondary variables through their effects on geometrical overlap and relative horizon-entry times.

This work provides the first fully nonlinear numerical-relativity study of
PBH formation from spatially distinct neighbouring curvature fluctuations,
directly resolving collective collapse and PBH multiplicity. Overall, our results show that the spatial environment of large curvature peaks must be accounted for when determining both the conditions for collapse and the resulting PBH multiplicity. The associated collective dynamics can modify the mapping from primordial curvature statistics to PBH abundances and spatial distributions, with potential consequences for PBH clustering, binary formation, and gravitational-wave signals from formation-era PBH mergers. As a next step, we plan to extend the dynamical analysis and the proposed threshold framework to more complex multi-peak geometries and assess their implications for PBH abundance and spatial distribution.

\section*{Acknowledgments}
We thank Chul-Moon Yoo for clarifications regarding the modification of the boundary conditions and window function in the \texttt{COSMOS} code. We acknowledge support from the APCTP Junior Research Group Leader program and JSPS KAKENHI Grant Number 26K17141. 


\bibliographystyle{JHEP}
\bibliography{bibfile}

\clearpage
\appendix

\section{Appendix: technical details}

\subsection{Hawking compactness on the initial slice}
\label{app:hawking_derivation}

For completeness, we derive the nonspherical compactness used in the main text.
We adopt signature $(-,+,+,+)$ and let $u^\mu$ be the future-directed unit
normal to the initial spatial slice. With
$K_{ij}=-(1/2){\cal L}_u\gamma_{ij}$, an expanding flat-FLRW background
has $K_{ij}^{\rm FLRW}=-H\gamma_{ij}$, where $H$ is the background Hubble
rate on the initial slice. For a closed two-surface $S$, let $s^i$ be its
outward unit normal within the spatial slice,
$q_{ij}=\gamma_{ij}-s_i s_j$ the induced metric,
$H_S=D_i s^i$ its mean curvature, and $P=q^{ij}K_{ij}$. Choosing the
future-directed null normals
\begin{equation}
 \ell^\mu=u^\mu+s^\mu,
 \qquad
 k^\mu=u^\mu-s^\mu,
 \qquad
 \ell\!\cdot\!k=-2,
\end{equation}
gives $\theta_+=H_S-P$, $\theta_-=-H_S-P$, and hence
$\theta_+\theta_-=P^2-H_S^2$. The Hawking quasi-local energy is therefore
\cite{Hawking:1968}
\begin{equation}
\begin{aligned}
 E_{\rm H}[S]
 &=\frac{R_A}{2}\left[
 1+\frac{1}{16\pi}\oint_S(P^2-H_S^2)\,dA
 \right],\\
 R_A&=\sqrt{\frac{A[S]}{4\pi}}.
\end{aligned}
\label{eq:end_EH}
\end{equation}
For a round sphere in flat FLRW, $P=-2H$ and $H_S=2/R_A$, so that
\begin{equation}
 E_{\rm H}^{\rm FLRW}(R_A)=\frac{H^2R_A^3}{2}.
\end{equation}
We define the equal-area FLRW-referenced compactness by
\begin{equation}
 {\cal C}_{\rm H}[S]\equiv\frac{2}{R_A}
 \left[E_{\rm H}[S]-E_{\rm H}^{\rm FLRW}(R_A)\right].
\label{eq:end_CH_general}
\end{equation}
The reference is a round flat-FLRW sphere at the same background time and
with the same area radius $R_A$. This prescription compares quasi-local
energies and does not sphericalize the physical curvature field.

To evaluate Eq.~\eqref{eq:end_CH_general} on the initial superhorizon slice,
we represent the growing mode in comoving slicing as
\begin{equation}
 \gamma_{ij}
 =a^2\Psi^4(\bm x)\left[\delta_{ij}+O(\epsilon^2)\right],
 \qquad
 \Psi=e^{\zeta/2},
 \qquad
 \epsilon\sim (a  H r_{m,\min})^{-1}\ll1.
\label{eq:end_metric}
\end{equation}
On a coordinate sphere
$S(\bm c,r):|\bm x-\bm c|=r$, define
\begin{equation}
 \bm n=\frac{\bm x-\bm c}{r},
 \qquad
 \zeta_n=n^i\partial_i\zeta,
 \qquad
 \langle F\rangle_\Omega
 \equiv
 \frac{1}{4\pi}\int F(\bm n)\,d\Omega .
\end{equation}
Here $\langle\cdots\rangle_\Omega$ denotes the ordinary angular average on the coordinate sphere, not an average weighted by its physical area. At the
order required below,
\begin{equation}
 dA=a^2e^{2\zeta}r^2\,d\Omega,
 \qquad
 H_S=\frac{2e^{-\zeta}}{ar}(1+r\zeta_n),
\label{eq:end_surface_geometry}
\end{equation}
and therefore
\begin{equation}
 \frac{1}{16\pi}\oint_S H_S^2\,dA
 =\left\langle(1+r\zeta_n)^2\right\rangle_{\Omega}.
\label{eq:end_Hpiece}
\end{equation}

The extrinsic-curvature contribution is fixed by the growing-mode gradient
expansion \cite{Harada:2015yda}. We write
\begin{equation}
 K=-3H(1+\kappa),
 \qquad
 K_{ij}=A_{ij}+\frac{K}{3}\gamma_{ij},
 \qquad
 A_{ij}=a^2\Psi^4\widetilde A_{ij},
\end{equation}
where $A_{ij}$ is traceless. For a perfect fluid satisfying $p=w\rho$, the
comoving-slicing growing-mode solution gives
\begin{align}
 \kappa
 &=-\frac{{\cal Q}}{5+3w}(aH)^{-2}+O(\epsilon^4),\\
 \widetilde A_{ij}
 &=\frac{2}{5+3w}\,p_{ij}H(aH)^{-2}+O(\epsilon^4H),
\label{eq:end_gradientK}
\end{align}
where
\begin{align}
 {\cal Q}&=-\frac{4}{3}\Psi^{-5}\nabla^2\Psi,\\
 p_{ij}
 &=\Psi^{-4}\Bigg[
 -\frac{2}{\Psi}
 \left(\partial_i\partial_j\Psi
 -\frac{1}{3}\delta_{ij}\nabla^2\Psi\right)
 \nonumber\\[-1mm]
 &\hspace{1.4cm}
 +\frac{6}{\Psi^2}
 \left(\partial_i\Psi\partial_j\Psi
 -\frac{1}{3}\delta_{ij}|\bm\nabla\Psi|^2\right)
 \Bigg].
\label{eq:end_pij}
\end{align}
Here spatial derivatives and $\nabla^2$ are taken with respect to the flat
conformal metric $\delta_{ij}$. Defining $A_{ss}=A_{ij}s^i s^j$, one has
\begin{equation}
 A_{ss}=\widetilde A_{ij}n^i n^j+O(\epsilon^4H)
\end{equation}
at the required order. Since
\begin{equation}
 P=\frac{2K}{3}-A_{ss},
\end{equation}
its square becomes
\begin{equation}
 P^2
 =4H^2+8H^2\kappa+4HA_{ss}
 +O(\epsilon^4H^2).
\end{equation}
The homogeneous contribution satisfies
\begin{equation}
 \frac{1}{16\pi}\oint_S4H^2\,dA=H^2R_A^2
\end{equation}
and therefore cancels the equal-area FLRW subtraction in
Eq.~\eqref{eq:end_CH_general}. The remaining contribution is
\begin{equation}
\begin{aligned}
 &\frac{1}{16\pi}\oint_S
 \left(8H^2\kappa+4HA_{ss}\right)\,dA\\
 &\qquad=
 \frac{2r^2}{5+3w}
 \left\langle
 \nabla^2\zeta-\zeta_{nn}+\zeta_n^2
 \right\rangle_{\Omega},
\end{aligned}
\label{eq:end_Kpiece}
\end{equation}
where $\zeta_{nn}=n^i n^j\partial_i\partial_j\zeta$. Using the
decomposition
\begin{equation}
 \nabla^2\zeta
 =\zeta_{nn}+\frac{2}{r}\zeta_n
 +\frac{1}{r^2}\Delta_\Omega\zeta
\end{equation}
and $\langle\Delta_\Omega\zeta\rangle_\Omega=0$, we obtain
\begin{equation}
\begin{aligned}
 {\cal C}_{\rm H}^{\rm co}(\bm c,r)
 &=\frac{3(1+w)}{5+3w}
 \left[
 1-\left\langle(1+r\zeta_n)^2\right\rangle_{\Omega}
 \right]\\
 &\quad+O(\epsilon^2).
\end{aligned}
\label{eq:end_CH_reduced}
\end{equation}
The superscript ``co'' indicates that this is the leading long-wavelength
representation in comoving slicing; it is suppressed in the main text. For
radiation domination, $w=1/3$ and the prefactor reduces to $2/3$.

In spherical symmetry, $\zeta_n=\zeta'(r)$ and
Eq.~\eqref{eq:end_CH_reduced} reduces to
\begin{equation}
 {\cal C}_{\rm H}^{\rm co}(r)
 =\frac{3(1+w)}{5+3w}
 \left[1-(1+r\zeta')^2\right]
 +O(\epsilon^2).
\label{eq:end_spherical}
\end{equation}
At the level of the exact geometrical definition,
$E_{\rm H}=M_{\rm MS}$ for a spherical surface. Consequently,
\begin{equation}
 {\cal C}_{\rm H}(r)
 =\frac{2}{R_A}
 \left[
 M_{\rm MS}(r)-M_{\rm MS}^{\rm FLRW}(R_A)
 \right],
\end{equation}
so Eq.~\eqref{eq:end_spherical} is the long-wavelength form of the standard
background-subtracted Misner--Sharp compaction function in our convention
$2\Delta M/R_A$ \cite{Shibata:1999zs,Harada:2015yda}.

Finally, define the curvature field obtained by sphericalizing about the
same trial centre,
\begin{equation}
 \bar\zeta_c(r)
 \equiv
 \left\langle
 \zeta(\bm c+r\bm n)
 \right\rangle_\Omega,
 \qquad
 \bar\zeta_c'(r)
 =
 \left\langle\zeta_n\right\rangle_\Omega .
\end{equation}
For radiation domination, the linear compaction associated with this
sphericalized profile is
\begin{equation}
 \mathcal C_\ell(\bm c,r)
 \equiv
 -\frac{4}{3}r\,\bar\zeta_c'(r)
 =
 -\frac{4}{3}r
 \left\langle\zeta_n\right\rangle_\Omega .
 \label{eq:end_Cl}
\end{equation}
The corresponding nonlinear sphericalized compactness is
\begin{equation}
 \begin{aligned}
 \mathcal C_{\rm sph}
 &=
 \frac{2}{3}
 \left[
  1-
  \left(
   1+r\left\langle\zeta_n\right\rangle_\Omega
  \right)^2
 \right]
 +O(\epsilon^2)
 \\
 &=
 \mathcal C_\ell
 -\frac{3}{8}\mathcal C_\ell^2
 +O(\epsilon^2).
 \end{aligned}
\end{equation}
Using
\[
 \left\langle\zeta_n^2\right\rangle_\Omega
 =
 \left\langle\zeta_n\right\rangle_\Omega^2
 +
 {\rm Var}_\Omega(\zeta_n),
\]
the full nonspherical Hawking compactness becomes
\begin{equation}
 \mathcal C_{\rm H}^{\rm co}
 =
 \mathcal C_\ell
 -\frac{3}{8}\mathcal C_\ell^2
 -\frac{2}{3}r^2
 {\rm Var}_\Omega(\zeta_n)
 +O(\epsilon^2),
 \label{eq:end_variance}
\end{equation}
where
\[
 {\rm Var}_\Omega(\zeta_n)
 \equiv
 \left\langle\zeta_n^2\right\rangle_\Omega
 -
 \left\langle\zeta_n\right\rangle_\Omega^2 .
\]

For a spherical profile, $\mathcal C_\ell=4/3$ corresponds to
$1+r\zeta'=0$, the boundary between the Type-I and Type-II branches of
the areal radius. For a nonspherical profile, however, $\mathcal C_\ell=4/3$ imposes only
$\langle1+r\zeta_n\rangle_\Omega=0$ and does not imply that
$1+r\zeta_n$ vanishes pointwise. We therefore use $\mathcal C_\ell$ as a Type-II-aware branch diagnostic, rather than as a new pointwise
geometrical classification. Equation~\eqref{eq:end_CH_general} is geometrical for a specified
two-surface, whereas Eq.~\eqref{eq:end_CH_reduced} is its comoving-slicing
long-wavelength representation on the chosen family of coordinate spheres. 

\subsection{Numerical implementation}

The initial data are evolved with the \texttt{COSMOS} code \cite{Yoo:2026itl}, originally translated into C++ from the \texttt{SACRA} numerical-relativity code \cite{Yamamoto:2008js}. The code evolves the Einstein field equations using the BSSN formulation
\cite{Shibata:1995we,Baumgarte:1998te}, coupled to a perfect fluid with equation of state $p=w\rho$, where $w=1/3$. The relativistic
hydrodynamic equations are written in flux-conservative $3+1$ form and
evolved using a MUSCL high-resolution central scheme
\cite{Kurganov:2000,Shibata:2005jv}. The hydrodynamic variables and evolution equations used here are summarized in Appendix E of
Ref.~\cite{Escriva:2024lmm}. Reflection
symmetries across $y=0$ and $z=0$ reduce the computational domain.

The numerical setup is the same as that used in Ref.~\cite{Escriva:2024lmm}, with a few modifications to accurately handle the present configurations and speed up the simulations. The base grid uses a double-focused coordinate map along the peak axis,
which increases the resolution around the individual perturbations. This is
particularly important when the fluctuations collapse separately, since
accurately resolving PBH formation requires high resolution in regions located
away from the coordinate origin. We denote the uniform computational coordinates by
$(\xi_x,\xi_y,\xi_z)$ and the corresponding comoving coordinates used
throughout the main text by $(x,y,z)$. Along the peak axis we use
\begin{equation}
x(\xi_x)=\mathcal N
\left[
\xi_x
-
A_\eta \sigma_g \sqrt{\frac{\pi}{2}}
\left\{
\operatorname{erf}
\left(
\frac{\xi_x-\xi_0}{\sqrt{2}\sigma_g}
\right)
+
\operatorname{erf}
\left(
\frac{\xi_x+\xi_0}{\sqrt{2}\sigma_g}
\right)
\right\}
\right],
\end{equation}
where
\begin{equation}
A_\eta=\frac{\eta}{1+\eta},
\qquad
x(\xi_0)=s,
\qquad
x(\pm L_{\rm box})=\pm L_{\rm box},
\end{equation}
and
\begin{equation}
r_{m,\max}\equiv\max(r_{m,1},r_{m,2}),
\qquad
\sigma_g=\frac{r_{m,\max}}{\sqrt{2}}.
\end{equation}
Here $\mathcal N$ is a normalization factor chosen so that the numerical
boundary remains at $x(\pm L_{\rm box})=\pm L_{\rm box}$.
Uniform sampling in $\xi_x$ therefore generates nonuniform sampling in
the comoving coordinate $x$, with enhanced resolution around the two
perturbations. The transverse-coordinate treatment is left unchanged.
Since the fixed mesh-refinement levels remain centred on the origin, the
double-focused base grid provides the principal resolution enhancement
around the off-centred collapsing regions.

%
%

To impose an exactly homogeneous geometry at the outer numerical
boundary, the physical profile is multiplied by the smooth radial window
\begin{equation}
 \zeta_{\rm num}(\bm x)=W(r)\zeta(\bm x),
\qquad
r=(x^2+y^2+z^2)^{1/2},
\end{equation}
with
\begin{equation}
\begin{aligned}
W(r)&=
\begin{cases}
1,& r\le r_{\rm in},\\
1-[1-u(r)^6]^6,& r_{\rm in}<r<L_{\rm box},\\
0,& r\ge L_{\rm box},
\end{cases}\\[1mm]
u(r)&=\frac{L_{\rm box}-r}{L_{\rm box}-r_{\rm in}}.
\end{aligned}
\end{equation}
We take
\begin{equation}
 r_{\rm in}=L_{\rm box}
 -\sqrt{\frac32}\,r_{m,\max}.
 \label{eq:supp_rin}
\end{equation}
with $L_{\rm box}=1.8L$ and require
\begin{equation}
 e^{-\left[(r_{\rm in}-s)/r_{m,\max}\right]^2}<10^{-2}
 \label{eq:supp_boundary_suppression}
\end{equation}
so that the window changes only the asymptotic tail.

The initial conformal factor is
\begin{equation}
 \Psi(\bm x)=e^{\zeta_{\rm num}(\bm x)/2}.
\end{equation}
For notational simplicity, in the remainder of this subsection we drop
the subscript ``num'' and denote the windowed curvature profile
$\zeta_{\rm num}$ simply by $\zeta$. The numerical initial data are
constructed using the constant-mean-curvature long-wavelength solution,
with $K=-3H$ and hence $\kappa=0$, whereas the compactness diagnostic in
Sec.~A.1 is expressed in comoving slicing. These gauge-related
long-wavelength descriptions share the same leading curvature profile
$\zeta$, while their density contrasts satisfy
\[
\delta_\rho^{\rm co}
=
\frac{3(1+w)}{5+3w}\delta_\rho^{\rm CMC}
+\mathcal{O}(\epsilon^4).
\]
For radiation domination, this relation becomes
$\delta_\rho^{\rm co}=(2/3)\delta_\rho^{\rm CMC}
+\mathcal{O}(\epsilon^4)$ \cite{Harada:2015yda}. Accordingly, the CMC
density contrast used in the numerical evolution,
$\delta_\rho\equiv(\rho-\rho_b)/\rho_b$, reads, at leading nontrivial
order,
\begin{equation}
 \delta_\rho
 =-\frac{4}{3(aH)^2}\Psi^{-5}\nabla^2\Psi
 +\mathcal{O}(\epsilon^4),
\end{equation}
which in terms of $\zeta$ becomes
\begin{equation}
 \delta_\rho
 =-\frac{4}{3(aH)^2}e^{-2\zeta}
 \left[
 \frac12\nabla^2\zeta+\frac14(\bm\nabla\zeta)^2
 \right]
 +\mathcal{O}(\epsilon^4).
\end{equation}
For the simulations we typically use
$H_{\rm ini}L=50$, $\eta=10$, $a_{\rm ini}=1$, and approximately
$80$ grid points in each spatial direction. As a resolution-robustness check, we repeated selected simulations with
approximately $130$ base-grid points in each spatial direction, keeping
the remaining numerical setup unchanged. All repeated configurations
retained the same classification as non-forming, single-PBH, or
double-PBH outcomes, including the same apparent-horizon multiplicity.

\begin{figure}[t]
\centering
\includegraphics[width=0.5\textwidth]{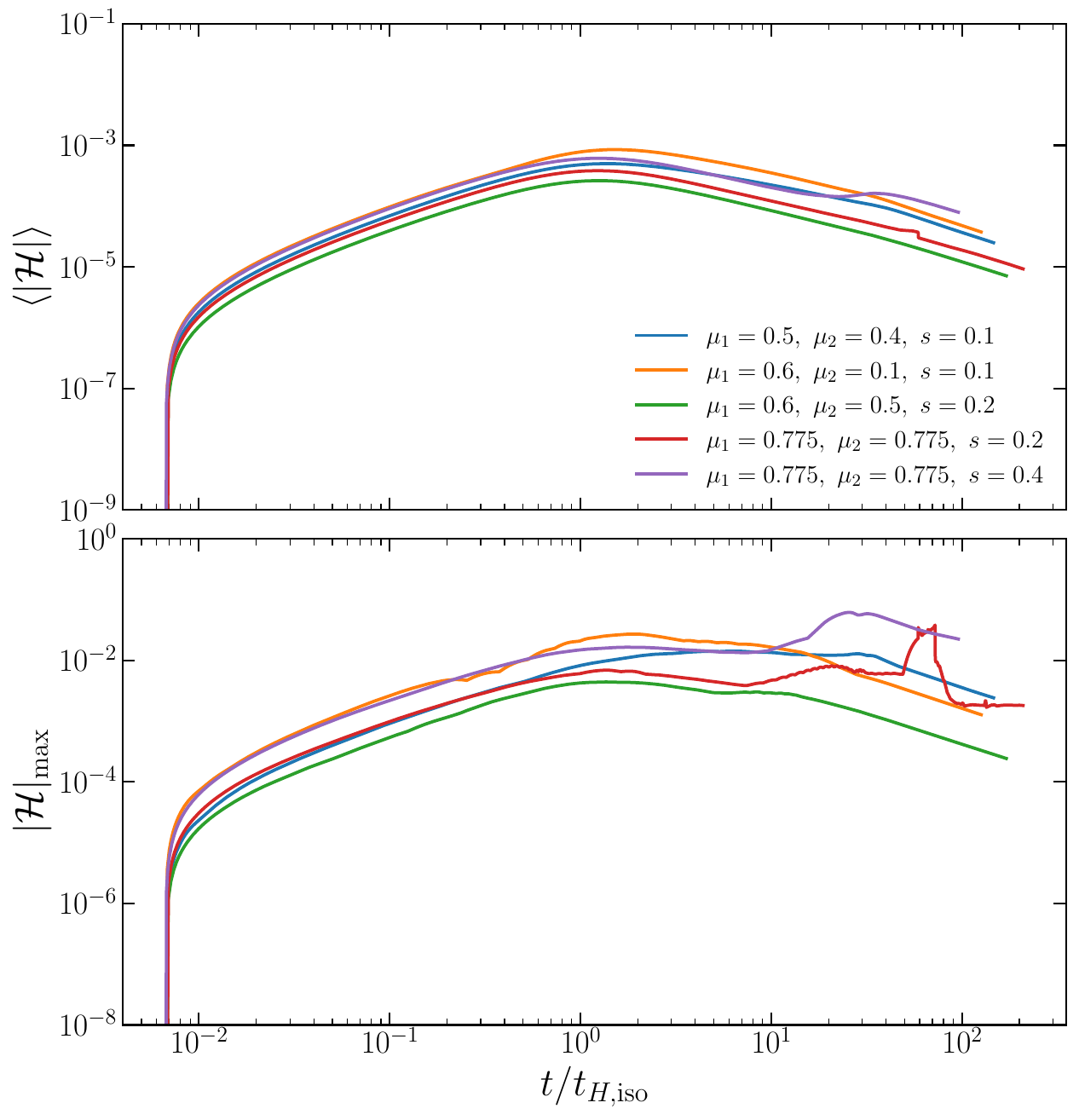}
\caption{Evolution of the Hamiltonian-constraint violation for representative
configurations. The upper panel shows the volume-averaged absolute value,
$\langle |\mathcal{H}| \rangle$, while the lower panel shows the maximum
absolute value, $|\mathcal{H}|_{\rm max}$, as functions of time normalized
by the horizon-entry time of a single fluctuation, $t_{H,\rm iso}$.}
\label{fig:prl_monitor}
\end{figure}

PBH formation
is monitored through trapped regions and apparent horizons, while the
Hamiltonian and momentum constraints are monitored throughout the
evolution. An example is shown in Fig.~\ref{fig:prl_monitor}. The different
curves correspond to the configurations indicated in the legend, spanning
both forming and non-forming cases. The constraint violation remains small
throughout the evolutions, including during the strongly nonlinear stages
associated with PBH formation.

\begin{figure}[t]
\centering
\includegraphics[width=0.5\textwidth]{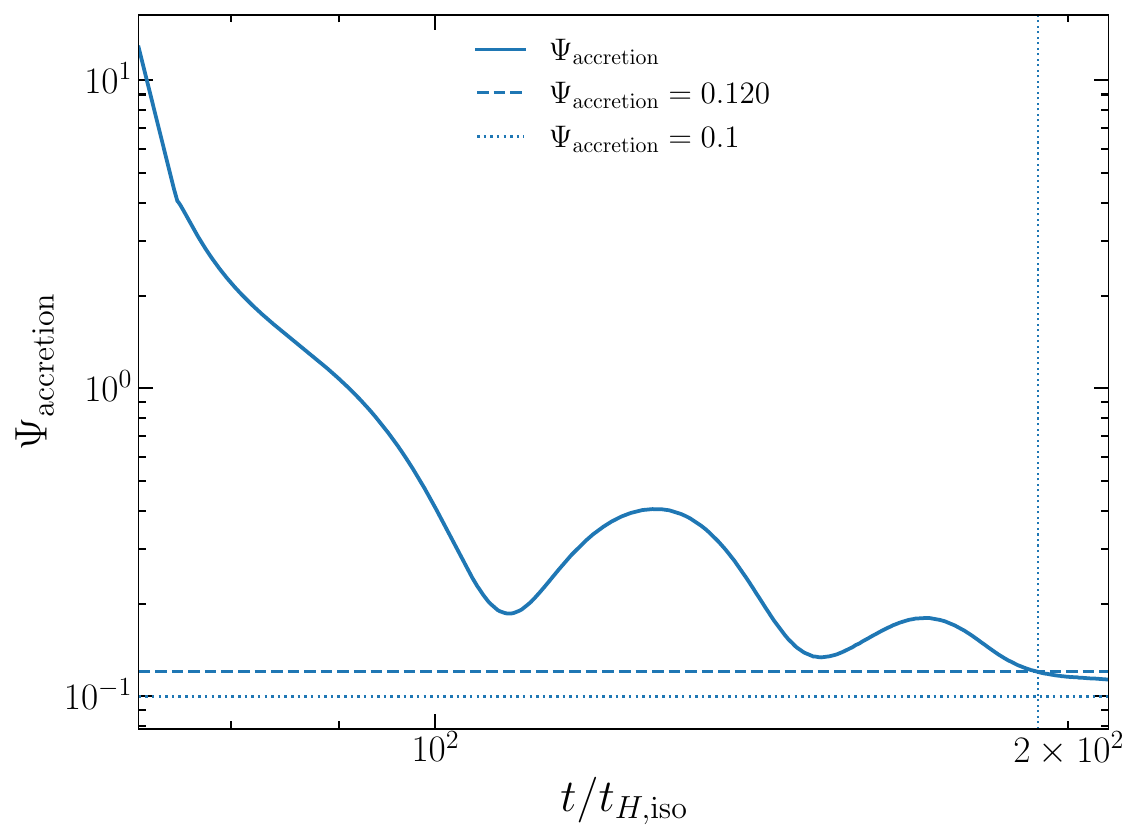}
\caption{
Evolution of the accretion diagnostic
$\Psi_{\rm accretion}$ for the collective-collapse configuration shown in Fig.~\ref{fig:pbh_mass_evolution}. The late-time fitting interval is selected from the final contiguous region
satisfying $\Psi_{\rm accretion} \leq 0.12$.
The dashed horizontal line indicates $\Psi_{\rm accretion}=0.12$, while the dotted horizontal
line shows the reference value $\Psi_{\rm accretion}=0.1$.
The vertical dotted line marks the beginning of the fitting interval.
Time is normalized by the horizon-crossing time $t_{H,\rm iso}$ associated
with the characteristic scale of a single isolated Gaussian component.}
\label{fig:evolution_Psi}
\end{figure}

The asymptotic PBH mass shown in Fig.~\ref{fig:pbh_mass_evolution} is estimated by fitting the
late-time evolution of the apparent-horizon mass to the
Zel'dovich--Novikov accretion model, following the procedure of
Refs.~\cite{Escriva:2019nsa,Escriva:2025eqc}. In a flat FLRW background, the accretion
rate is approximated by
\begin{equation}
    \frac{dM_{\rm PBH}}{dt}
    =
    4\pi F R_{\rm PBH}^{\,2}\rho_b(t),
\end{equation}
where \(F\) is an effective accretion parameter. Using
\(R_{\rm PBH}=2M_{\rm PBH}\) and \(H=1/(2t)\) during radiation
domination, this gives
\begin{equation}
    M_{\rm PBH}(t)
    =
    \left[
        \frac{1}{M_\infty}
        +\frac{3F}{2t}
    \right]^{-1},
\label{eq:pbh_accretion_fit}
\end{equation}
where \(M_\infty\equiv M_{\rm PBH}(t\rightarrow\infty)\) is the
asymptotic PBH mass. We determine \(M_\infty\) and \(F\) by fitting
Eq.~\eqref{eq:pbh_accretion_fit} directly to the numerical
apparent-horizon mass at sufficiently late times, when the accretion
flow approaches the asymptotic regime. To identify this regime, we monitor the accretion diagnostic
\cite{Escriva:2019nsa}
\begin{equation}
    \Psi_{\rm accretion} \equiv
    \frac{\dot M_{\rm PBH}}{H M_{\rm PBH}},
\end{equation}
and perform the fit over the final contiguous interval satisfying
\(\Psi_{\rm accretion}\lesssim 0.12\), as shown in Fig.~\ref{fig:evolution_Psi}. Varying the
late-time selection around this value gives a stable estimate of
\(M_\infty\). For the collective-collapse configuration shown in
Fig.~\ref{fig:prl_single_dynamics1}, we obtain
\(M_\infty/M_{H,\rm iso}\simeq 2.85\).

\begin{figure*}[t]
\centering
\includegraphics[width=1.0\textwidth]{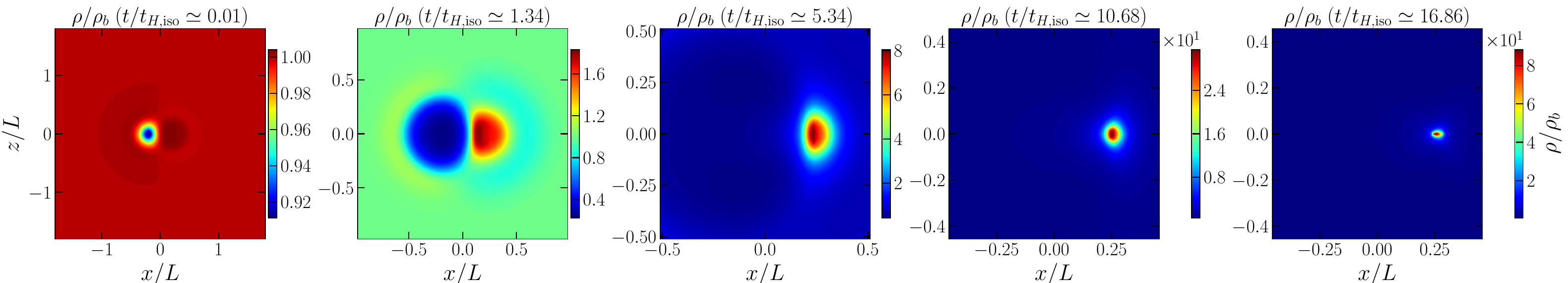}
\includegraphics[width=0.31\textwidth]{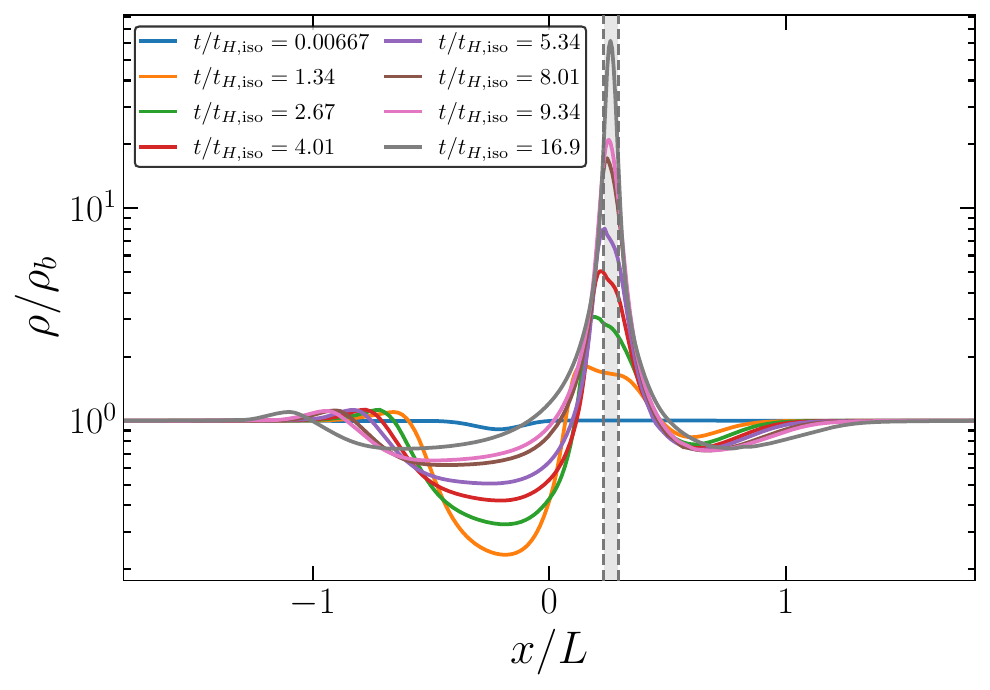}
\includegraphics[width=0.31\textwidth]{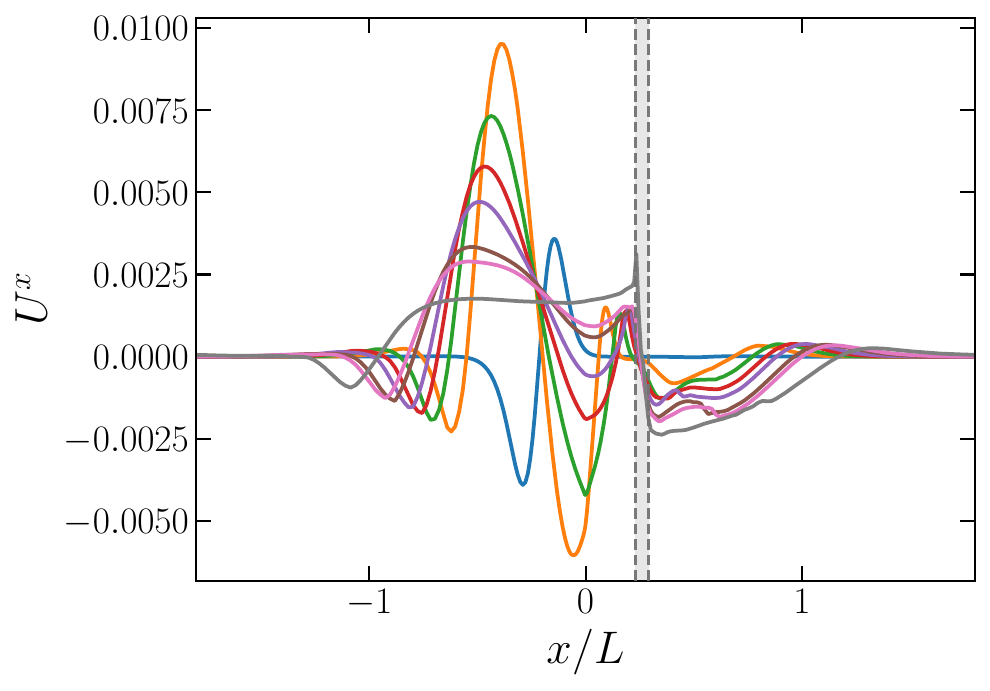}
\includegraphics[width=0.31\textwidth]{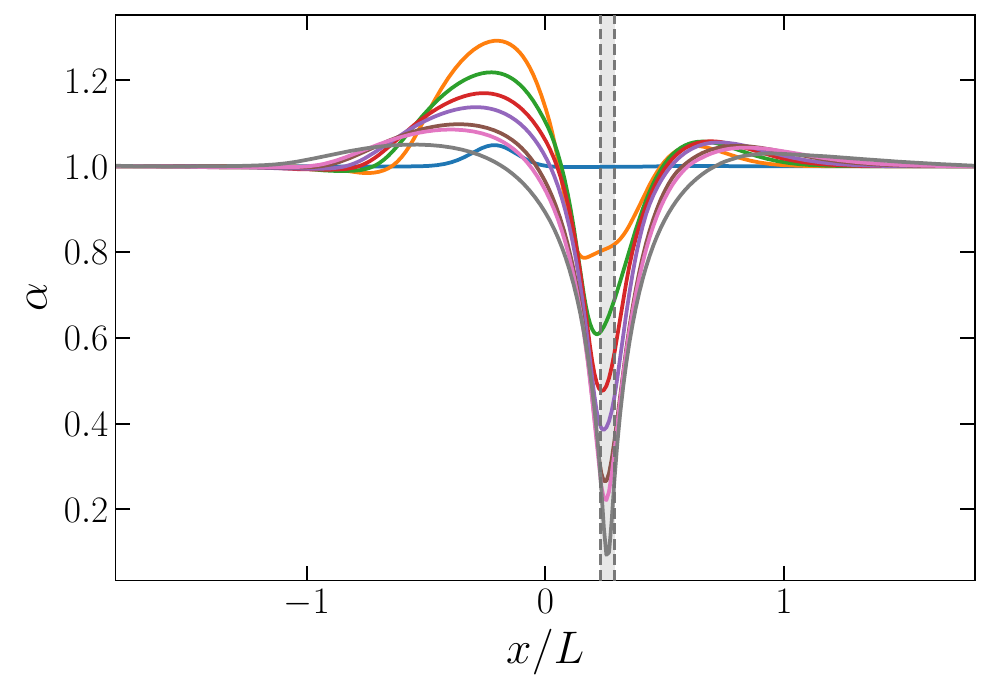}
\caption{Same dynamical diagnostics as in
Fig.~\ref{fig:prl_single_dynamics1}, for a single-PBH configuration with
$(\mu_1,\mu_2,s)=(1.2,-0.8,0.2)$. The grey shaded band marks the first detected local apparent horizon, with coordinate extent $0.2299\lesssim x/L\lesssim0.2914$, appearing at $t/t_{H,\mathrm{iso}}\simeq16.86$.}
\label{fig:prl_single_dynamics55}
\end{figure*}

\begin{figure*}[t]
\centering
\includegraphics[width=0.8\textwidth]
{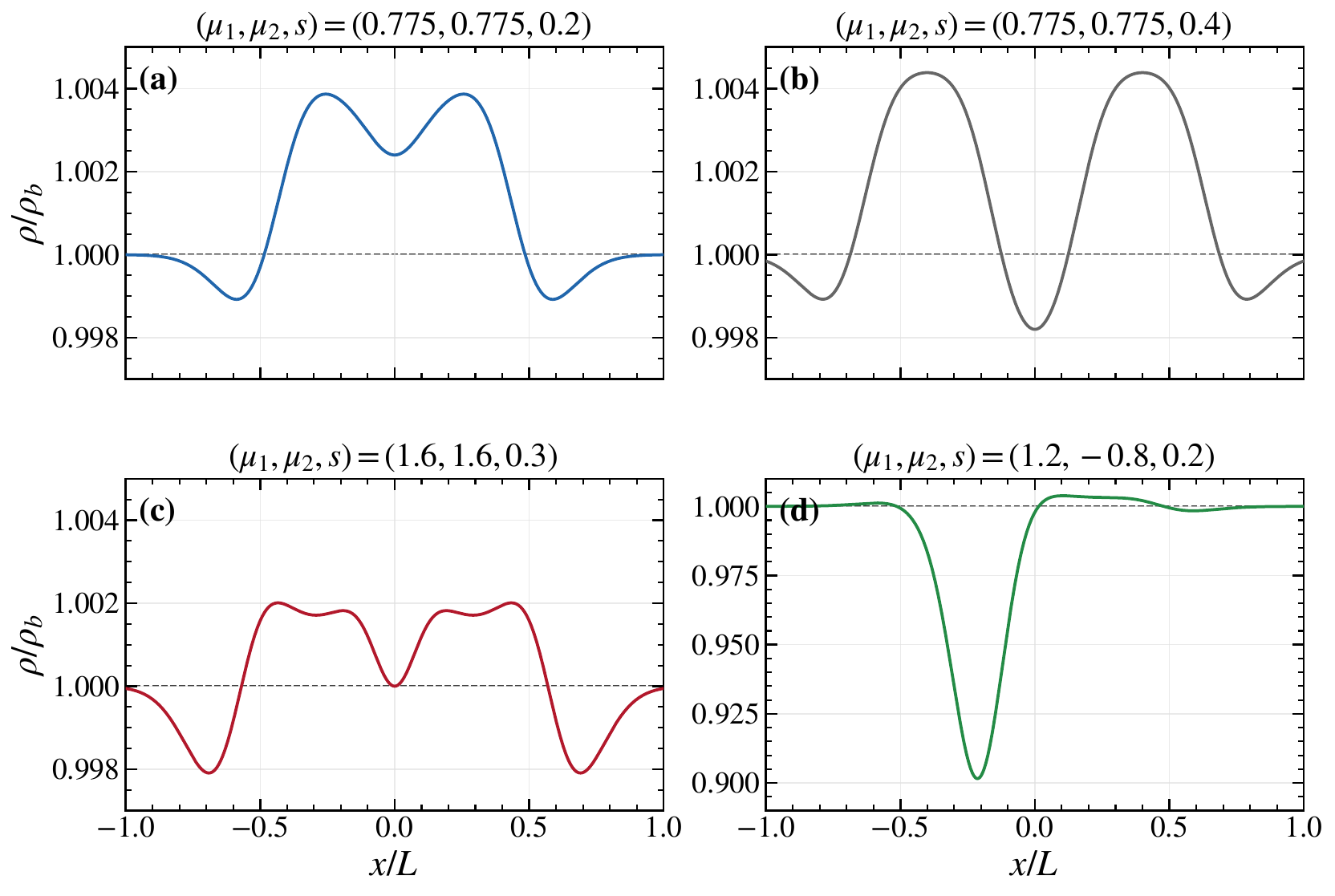}
\caption{Initial axial density profiles for the four dynamical examples discussed in the text. Shown is $\rho/\rho_b$ along $y=z=0$ on the initial superhorizon slice. Panels (a)--(d) correspond, respectively, to the collective single-PBH, separated non-forming, double-PBH, and mixed-sign single-PBH configurations. All cases have $r_{m,1}/L=r_{m,2}/L=\sqrt{6}/10$ and $H_{\rm ini}L=50$. The horizontal dashed line marks the homogeneous FLRW background, $\rho/\rho_b=1$.}
\label{fig:supp_initial_density_profiles}
\end{figure*}

In Fig.~\ref{fig:prl_single_dynamics55}, we present an additional example of the nonlinear dynamics, corresponding to a mixed-sign pair of neighbouring fluctuations. At early times, the negative-amplitude component generates a pronounced underdense region on the negative-$x$ side, while the positive component evolves into an increasingly compact overdensity around $x/L\simeq0.25$. The axial velocity develops the sign change characteristic of convergent flow towards this overdensity, and the lapse exhibits a progressively deeper and more localized minimum at the same position. By contrast, the underdense component disperses and never develops a trapped surface. Consequently, only the positive peak forms a PBH, with the first local apparent horizon detected at $t/t_{H,\mathrm{iso}}\simeq16.86$. The initial axial density profiles of the four representative dynamical configurations are compared in Fig.~\ref{fig:supp_initial_density_profiles}.

\subsection{Additional basin-resolved diagnostics}

The global Hawking formation strength $\mathcal K_{\rm form}$, the nonlinear weaker-local strength $\mathcal K_{\mathrm{loc}}$, and the linear branch strengths $\mathcal K_{\ell,\mathrm{loc}}$ and $\mathcal K_{\ell,C}$ are defined in the main text, together with the
normalized variables $\nu_{\rm loc}$ and $\mathcal D_\ell$. Across the numerical sample, the largest
non-forming value of $\mathcal K_{\rm form}$ is $\sim 0.558$, whereas the smallest forming value is $\sim 0.555$. Their narrow overlap motivates the conservative formation-transition interval quoted in the main text rather
than a sharp universal threshold.

To further visualize the compactness landscapes, we introduce an auxiliary
common-enclosing construction. For a bimodal curvature profile with maxima
at $x_{\zeta,L}<x_{\zeta,R}$, the smallest radius of a sphere centred at
$x_c$ that encloses both maxima is
\begin{equation}
 r_{\rm enc}(x_c)
 =
 \max\!\left(
 |x_c-x_{\zeta,L}|,
 |x_c-x_{\zeta,R}|
 \right).
 \label{eq:supp_renc}
\end{equation}
The corresponding common-enclosing domain and compactness strength are
\begin{equation}
 \mathcal B_C
 =
 \{(x_c,r):r\geq r_{\rm enc}(x_c)\},
 \qquad
 \mathcal K_C
 =
 \max_{(x_c,r)\in\mathcal B_C}{\cal C}_H(x_c,r).
 \label{eq:supp_KC}
\end{equation}
The quantity $\mathcal K_C$ characterizes the nonlinear common-enclosing
portion of the Hawking-compactness landscape. The same domain
$\mathcal B_C$ is used to define the linear common-enclosing strength
$\mathcal K_{\ell,C}$ in Eq.~\eqref{eq:prl_Kell_branches}; therefore,
the linear comparison introduces no additional geometrical partition.

The local domains $\mathcal B_L$ and $\mathcal B_R$ are defined in the
main text, while the common-enclosing domain $\mathcal B_C$ is defined
in Eq.~\eqref{eq:supp_KC}. These domains
intentionally do not exhaust the $(x_c,r)$ plane. Spheres that cross the
saddle plane but do not yet enclose both curvature maxima are classified as
intermediate surfaces. They remain included in the unrestricted maximization
defining $\mathcal K_{\rm form}$ in Eq.~\eqref{eq:prl_Kform}, but are not
assigned to either a local or common-enclosing branch.

\begin{figure*}[t]
\centering
\includegraphics[width=0.99\textwidth]
{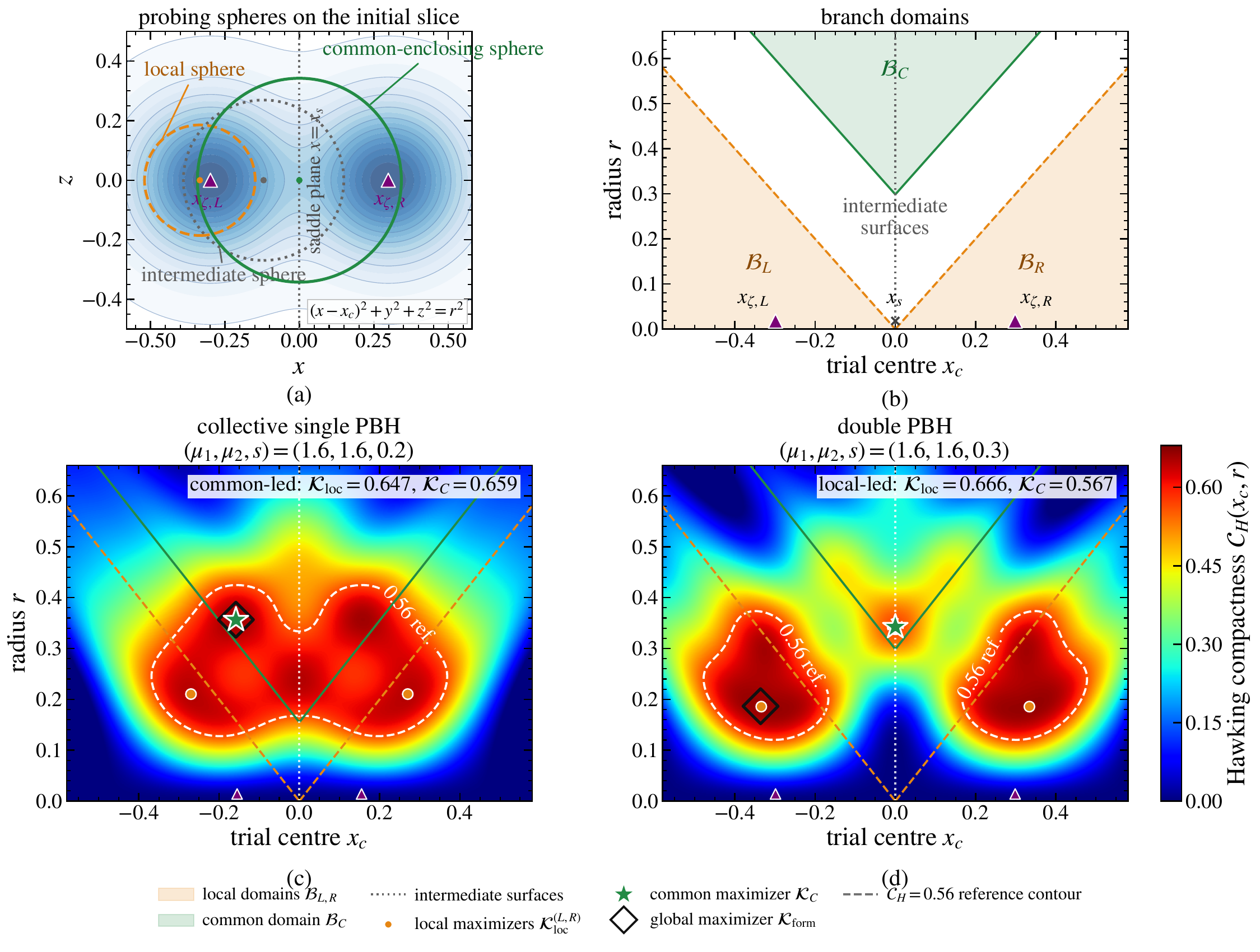}
\caption{Basin-resolved Hawking-compactness construction.
(a) Cross-section in the $y=0$ plane of representative coordinate probing
spheres for the profile used in panel (d). The orange dashed circle is a
left-local sphere, the green solid circle encloses both curvature maxima, and
the gray dotted circle illustrates an intermediate sphere that crosses the
saddle plane without enclosing both maxima.
(b) Corresponding domains in the trial-centre--radius plane. The orange
regions $\mathcal B_L$ and $\mathcal B_R$ contain spheres confined to one
side of the saddle plane, whereas the green region $\mathcal B_C$ contains
spheres enclosing both curvature maxima. The unshaded region represents
intermediate surfaces.
(c),(d) Full $\mathcal C_H(x_c,r)$ landscapes for a strong collective
single-PBH configuration and a strong double-PBH configuration,
respectively. Purple triangles at $r=0$ locate the curvature maxima, and
the vertical dotted line marks $x_s$, the location of the intervening
axial minimum defining the saddle plane. Orange circles mark
the local maximizers, the green star marks the auxiliary common-enclosing
maximizer, and the open black diamond marks the unrestricted global
maximizer. Marker overlap is physical:
$\mathcal K_{\rm form}\simeq\mathcal K_C$ in panel (c), whereas
$\mathcal K_{\rm form}\simeq\mathcal K_{\mathrm{loc}}$ in panel (d). The white dashed contour at $\mathcal C_H=0.56$ is only a broad formation
reference and is not by itself a multiplicity boundary. The same local and
common-enclosing domains are used to evaluate the linear branch strengths
$\mathcal K_{\ell,\mathrm{loc}}$ and $\mathcal K_{\ell,C}$ entering
Fig.~\ref{fig:prl_classification}.}
\label{fig:figS_hawking_basin_landscapes.pdf}
\end{figure*}

Figure~\ref{fig:figS_hawking_basin_landscapes.pdf} illustrates the common
geometrical construction in the $(x_c,r)$ plane. For the collective
single-PBH configuration $(\mu_1,\mu_2,s)=(1.6,1.6,0.2)$, the unrestricted
Hawking maximum lies on the common-enclosing branch, with
$\mathcal K_{\rm form}\simeq\mathcal K_C>
\mathcal K_{\mathrm{loc}}$.
The corresponding linear strengths are
$\mathcal K_{\ell,\mathrm{loc}}\simeq1.443$ and
$\mathcal K_{\ell,C}\simeq1.721$, giving
$\mathcal D_\ell\simeq-0.356$.

For the double-PBH configuration
$(\mu_1,\mu_2,s)=(1.6,1.6,0.3)$, the Hawking maximum instead lies on a local branch, with
$\mathcal K_{\rm form}\simeq\mathcal K_{\mathrm{loc}}>\mathcal K_C$. In this case,
$\mathcal K_{\ell,\mathrm{loc}}\simeq1.522$ and
$\mathcal K_{\ell,C}\simeq1.071$, giving
$\mathcal D_\ell\simeq0.577$. Both examples have
$\mathcal K_{\ell,\mathrm{loc}}>4/3$, so the nonlinear local
compactness is already near its Type-I/Type-II turnover and cannot order their strengths monotonically. By contrast, the signed linear branch contrast $\mathcal D_\ell$
distinguishes whether the weaker local branch or the common-enclosing branch is stronger. Together with $\nu_{\rm loc}$, it provides a clearer empirical separation of the multiplicity outcomes, although some overlap
remains near the transition.

\clearpage
\end{document}